\documentclass[12pt]{article}
\usepackage{graphicx}
\usepackage{epstopdf}
\usepackage{amssymb}
\usepackage{amsmath}
\usepackage{amsfonts}
\usepackage{amssymb}
\usepackage{dcolumn}
\usepackage[colorlinks = true, allcolors = blue]{hyperref}
\usepackage{subfig}
\usepackage{float}
\usepackage{amsbsy}
\usepackage[T1]{fontenc}
\newcommand{\doublespacing}{\let\CS=\@currsize\renewcommand{\baselinesstrech}
{2.0}\tiny\CS}
\begin{document}
\newcommand{\bd}{\begin{document}}
\newcommand{\ed}{\end{document}}
\newcommand{\bc}{\begin{center}}
\newcommand{\ec}{\end{center}}
\newcommand{\bfr}{\begin{flushright}}
\newcommand{\efr}{\end{flushright}}
\newcommand{\lt}{\left}
\newcommand{\rt}{\right}
\newcommand{\vs}{\vspace}
\newcommand{\hs}{\hspace}
\newcommand{\beq}{\begin{equation}}
\newcommand{\eeq}{\end{equation}}
\newcommand{\lb}{\linebreak}
\newcommand{\pb}{\pagebreak}
\newcommand{\mb}{\makebox}
\newcommand{\fb}{\framebox}
\newcommand{\mc}{\multicolumn}
\newcommand{\ben}{\begin{enumerate}}
\newcommand{\een}{\end{enumerate}}
\newcommand{\bit}{\begin{itemize}}
\newcommand{\eit}{\end{itemize}}
\newcommand{\oln}{\overline}
\newcommand{\un}{\underline}
\newcommand{\lefq}{\lefteqn}
\newcommand{\ba}{\begin{array}}
\newcommand{\ea}{\end{array}}
\newcommand{\beqa}{\begin{eqnarray}}
\newcommand{\eeqa}{\end{eqnarray}}
\newcommand{\beqas}{\begin{eqnarray*}}
\newcommand{\eeqas}{\end{eqnarray*}}
\newcommand{\bfg}{\begin{figure}}
\newcommand{\efg}{\end{figure}}
\newcommand{\bds}{\begin{displaymath}}
\newcommand{\eds}{\end{displaymath}}
\newcommand{\btb}{\begin{tabbing}}
\newcommand{\etb}{\end{tabbing}}
\newcommand{\para}{\parallel}
\newcommand{\pad}{\partial}
\newcommand{\nn}{\nonumber}
\newcommand{\la}{\leftarrow}
\newcommand{\ra}{\rightarrow}
\newcommand{\lgla}{\longleftarrow}
\newcommand{\lgra}{\longrightarrow}
\newcommand{\La}{\Leftarrow}\newcommand{\Ra}{\Rightarrow}
\newcommand{\Lra}{\Leftrightarrow}
\newcommand{\Lgla}{\Longleftarrow}
\newcommand{\Lgra}{\Longrightarrow}
\newcommand{\lan}{\langle}
\newcommand{\ran}{\rangle}
\renewcommand{\a}{\alpha}
\renewcommand{\b}{\beta}
\newcommand{\g}{\gamma}
\newcommand{\G}{\Gamma}
\renewcommand{\d}{\delta}
\newcommand{\eps}{\epsilon}
\newcommand{\Th}{\Theta}
\newcommand{\s}{\sigma}
\newcommand{\lam}{\lambda}
\newcommand{\D}{\Delta}
\newcommand{\ds}{\displaystyle}
\newcommand{\vare}{E}
\newcommand{\pr}{\prime}
\newcommand{\ro}{\rho}
\newcommand{\nab}{\nabla}
\newcommand{\m}{\mu}
\newcommand{\n}{\nu}
\newcommand{\Sg}{\Sigma}
\newcommand{\p}{\pi}
\newcommand{\R}{I\!\!R}
\newcommand{\om}{\omega}
\newcommand{\Om}{\Omega}
\newcommand{\ovra}{\overrightarrow}
\newcommand{\ze}{\zeta}
\newcommand{\vart}{\vartheta}
\newcommand{\tri}{\triangle}
\newcommand{\f}{\frac}
\newcommand{\iny}{\infty}
\newcommand{\pro}{\propto}
\renewcommand{\arraystretch}{1.25}

\bc {\large\bf Higher order complex cubic quintic Ginzburg Landau equation
	: Chirped solitary waves} \ec

\vs{0.5cm}

\bc
{\it {\bf Naresh Saha} {\footnote{e-mail : saha.naresh92@gmail.com}}, {\bf Barnana Roy}{\footnote{e-mail : taturoy@gmail.com}} \\
Physics \& Applied Mathematics Unit,\\
Indian Statistical Institute,\\
Kolkata - 700108, India.\\
~~\\

{\bf Avinash Khare} {\footnote{e-mail : avinashkhare45@gmail.com}}\\
Department of Physics, \\
Sabitribai Phule Pune University,\\
 Pune - 411007, India.} \ec
\vs{1.0cm}

\bc {\large {\un{Abstract}}} \ec 
Propagation characteristics of the chirped dissipative solitary waves
are investigated within the framework of higher order complex cubic quintic
Ginzburg Landau equation. Potentially rich set of exact chirped dissipative 
pulses, such as, bright, dark, grey, antidark, kink, antikink are derived in 
the presence of the self steepening, self frequency shift and nonlinear 
gain/loss. The linear stability results are corroborated by the direct 
numerical simulations. The effect of the variation of model 
parameters on physical quantities like the speed, amplitude and chirping are 
explored.

\newpage
\section{Introduction}
Dissipative solitons \cite{akhme1} are a class of localized structures of the 
nonconservative nonlinear system that exists due to the balance between the
dispersion (diffraction) and a conservative nonlinear effect and also between 
the linear or the nonlinear loss and amplification. The concept of dissipative solitons is applicable to the solid 
state lasers \cite{moor} as well as to the fiber optics communication systems 
for the description of pulse propagation in fibres in the presence of linear and 
nonlinear gain and spectral filtering \cite{matsu}, pulse generation in fiber lasers with additive pulse mode-locking \cite{haus,moor}, or pulse 
propagation in hollow core photonic crystal fibers filled with resonant gases 
\cite{facao}. The basic mathematical model for the description of all these 
phenomena in optics is the complex cubic-quintic Ginzburg-Landau equation 
(CCQGLE). \cite{akhme661}.
Dissipative solitons have been observed 
experimentally in \cite{crespo}. Apart from supporting various localized 
solutions like stationary solitons, sources, sinks, moving solitons and fronts 
\cite{marq}, the CCQGLE also 
posseses pulsating, creeping, erupting and chaotic solitons \cite{akhme20}. 
Exact chirped soliton solutions e.g. fixed and arbitrary amplitude, flat top 
solitons as well as some special pulse solutions with sharp peak at the edges 
and a dip in the center etc. of CCQGLE is given in \cite{akhme10}. \\ 
For studies of ultrashort pulses in optical fibers, CCQGLE should 
include higher order terms like third order dispersion (TOD), fourth order dispersion (FOD) and nonlinear gradient terms like the self steepening (SS), self frequency shift (SFS), dispersion of the nonlnear refractive index and the 
nonlinear gain \cite{deiss1990}. It has been shown in \cite{latas1} that these higher order effects have marked
impact on the dynamics of the pulsating, creeping, erupting and chaotic 
solutions. Specifically, for some range of the parameter values, the periodic 
as well as the chaotic behavior of some of these pulses is eliminated and they 
are transformed into fixed shape solutions. The transition from stationary to 
the pulsating solutions of CCQGLE under the influence of the nonlinear gain, 
its saturation and higher order effects like the SS, TOD and intrapulse Raman scattering (IRS) has been analyzed in \cite{uzu1}. 
Impact of the SFS, SS and TOD on the soliton explosion has been studied in \cite{schalte}. The role of higher order nonlinear and dispersions on the periodic (non chaotic) and chaotic exploding 
solitons is investigated in \cite{cartes2016}. The effect of the nonlinear gradient terms on breathing dissipative solitons is 
studied in \cite{deiss1998}. It is shown that even small nonlinear gradient 
terms can make the soliton either to breathe periodically or chaotically on 
only one side or to spread rapidly. The influence of TOD, IRS and SS on the localized fixed shape soliton 
solution of CCQGLE is studied with the help of the soliton perturbation theory 
in \cite{hori14}. These perturbations are found to affect the speed but not the 
structural stability of the pulses which behave like solitons. Existence of several specific families of robust stationary and
oscillating bound states of solitons in CCQGLE with TOD is 
shown in \cite{saka37}. The effect of the nonlinear gradient terms on the 
localized states in CCQGLE is investigated in \cite{deiss1990}. It is found 
that these terms affect the speed of the states as well as cause an asymmetry. 
Modulational instability of the continuous wave solution to the higher order 
CCQGLE has been studied in \cite{tiofack2009}. An exact analytic front solution 
for CCQGLE with higher order terms is given in \cite{tian10} and the necessary 
stability condition is derived. \\
The influence of higher order dispersion on a chirped pulse 
oscillator has been studied in \cite{kalash13}. It is shown that while positive FOD enhances the oscillator stability, the odd 
negative higher order dispersions weaken the stability but broaden the spectrum. The influence of strong stimulated Raman scattering (SRS) on the 
formation of a complex of the bound dissipative soliton and Raman pulse of comparable energy is shown in \cite{bednya}. The chirped dissipative soliton stability under the influence of quantum noise and SRS is studied in \cite{kalash14}. Also the formation of a new kind of dissipative soliton namely, Raman soliton for the CCQGLE in the presence of SRS is reported. The latter significantly affects the energy scalability of the Raman soliton. Chirped lambert W-kink solitons, and optical shock type solitons are obtained in \cite{nisha2020} for CCQGLE in the presence of IRS. The nonlinear fiber laser has been modeled using CCQGLE including SS and IRS effects in \cite{belan6}. It is shown that above a minimum value of the IRS effect, it is possible to find two chirped solitary pulses for the laser system. The smaller one belongs to the dispersion managed regime 
whereas the larger one belongs to the so called similariton regime. Chirped soliton like pulse solution to the higher order 
extended Ginzburg Landau equation with the FOD is obtained in \cite{tian2005}. The existence and the stability conditions of the model parameters have been analyzed.\\
As it stands, the study of propagation dynamics of the chirped 
solitary wave solutions to CCQGLE including SS and SFS along with other nonlinear gradient terms, appears to be not that well studied in the literature. As CCQGLE is considered as one of the paradigms for the localized state formation in mode 
locked lasers \cite{schalte}, it is very important to know the range of values of the coefficients for which the CCQGLE has stable solutions. This can be dealt with by finding exact solutions and studying their stability. This has motivated us to find exact chirped solitary wave solutions 
to the CCQGLE with SS, SFS and other nonlinear gradient terms. We report a variety of dissipative pulses, namely, the bright, dark, grey, antidark, kink, and antikink (subject to some parametric conditions). The novelty of the present formulation lies in the existence of grey, antidark and antikink dissipative solitary waves. To the best of our knowledge, these have not been reported so far in the literature. The results of linear stability analysis have been corroborated by direct numerical simulation. It is found that the solitary waves have stable evolution up to a distance of approximately $1000$ units along the direction of propagation beyond which the instability in the propagation of all the solitary waves is seen. Also the effect of variation of the independent model parameters on the speed, amplitude and the chirping are explored.\\ 
\section{Analytical Procedure }
We consider the following generalized CCQGLE \cite{latas1} 
\begin{equation}\label{1}
		i\psi_{z}+\frac{p_r}{2}\psi_{tt}+ ip_i\psi_{tt} + |\psi|^2\psi + iq_i|\psi|^2\psi +(c_r+ic_i)|\psi|^4\psi=i(m_r+im_i)(|\psi|^2\psi)_{t}+i(n_r+in_i)|\psi|^2_{t}\psi
\end{equation}
where in the optical context, $\psi(z,t)$ is the normalized envelope of
the optical field, $z$ is the normalized propagation distance and $t$ is the 
retarded time in the frame moving with the pulse. $p_r=\pm 1$ is the group 
velocity dispersion coefficient, $p_i$ describes the spectral filtering, $c_r$ 
is a higher order correction term to the nonlinear refractive index (quintic 
Kerr effect), $q_i$ accounts for nonlinear gain [and (or) absorption 
processes], $c_i$ describes the saturable effects of the nonlinear gain [and 
(or absorption)], $m_r$ is associated with the SS \cite{agra,lam},
$m_i$ is associated with the frequency dependency of the nonlinear gain/loss, 
$n_i$ is the intrapulse Raman scattering which is responsible for the soliton 
SFS \cite{agra,carval87} while $n_r$ accounts for the delayed 
nonlinear gain/loss response.

We proceed to search for chirped solutions of Eqn.(1) in the form
\begin{equation} 
	\psi(z,t)=\rho(\xi)e^{i[\chi(\xi)-\kappa z]}
\end{equation}
where $\xi=t-uz$ is the traveling coordinate, $\rho$, $\chi$ are real 
functions of $\xi$ and $\kappa$ is a real constant. Here $u = \frac{1}{v}$, $v$
being the group velocity of the wave packet. The corresponding chirp is given 
by $\delta\omega(t,z) = - \frac{\partial}{\partial t}[\chi(\xi) - \kappa z] = -
\frac{d}{d\xi}\chi(\xi)$. Substitution of Eqn.(2) into Eqn.(1) followed by a 
separation of the real and the imaginary parts yield the following two coupled 
equations in $\rho$ and $\chi$.
\begin{equation}\label{2}
		u\chi^{'}\rho+\kappa\rho+\frac{p_r}{2}\rho^{''}-\frac{p_r}{2}\rho\chi^{'2}-p_i\rho\chi^{''}-2p_i\rho^{'}\chi^{'}+\rho^3+c_r\rho^5+m_r\chi^{'}\rho^3+(3m_i+2n_i)\rho^2\rho^{'}=0
\end{equation}
\begin{equation}\label{3}
		-u\rho^{'}+p_i\rho^{''}-p_i\rho\chi^{'2}+p_r\rho^{'}\chi^{'}+\frac{p_r}{2}\rho\chi^{''}+q_i\rho^3+c_i\rho^5-3m_r\rho^2\rho^{'}+m_i\rho^3\chi^{'}-2n_r\rho^2\rho^{'}=0
\end{equation}
To solve the coupled Eqns.(3) and (4) we choose
\begin{equation}
	\chi^{'}=\alpha\rho^2+\beta
\end{equation}
so that the frequency chirping is given by $\delta \omega(t,z) = 
- \chi^{'}(\xi) = -(\alpha \rho^2 + \beta)$ where $\alpha$ and $\beta$ are the 
nonlinear and constant chirp parameters respectively.
Using Eqn.(5) in Eqns.(3) and (4) yield
\begin{equation}\label{4}
	\begin{split}
		(-\alpha^2\frac{p_r}{2}+c_r+m_r\alpha)\rho^5+(\alpha u-\alpha\beta p_r+m_r\beta+1)\rho^3+(\beta u+\kappa-\frac{p_r}{2}\beta^2)\rho\\+\frac{p_r}{2}\rho^{''}
		+(-4\alpha p_i+3m_i+2n_i)\rho^2\rho^{'}+(-2\beta p_i)\rho^{'}=0
	\end{split}
\end{equation}
\begin{equation}\label{5}
	\begin{split}
		(-\alpha^2p_i+c_i+m_i\alpha)\rho^5+(-2\alpha\beta p_i+m_i\beta+q_i)\rho^3-p_i\beta^2\rho+p_i\rho^{''}\\
		+(2\alpha p_r-3m_r-2n_r)\rho^2\rho^{'}+(-u+\beta p_r)\rho^{'}=0
	\end{split}
\end{equation}
The following relations among the various model parameters
\begin{equation}\label{6}
	\begin{split}
		\alpha=\frac{3m_r+2n_r}{2p_r},~\beta=\frac{u}{p_r},~p_i=0,~q_i=-\frac{m_iu}{p_r}\\
		c_i=-\frac{m_i(3m_r+2n_r)}{2p_r},~~m_i=-\frac{2n_i}{3}
	\end{split}
\end{equation}

reduce Eqns.(6) and (7) into a single equation
\begin{equation}\label{8}
	\rho^{''}+a_1\rho^5+a_2\rho^3+a_3\rho=0,
\end{equation}
\begin{gather*}
	\text{where  }a_1=\frac{1}{4p_r^2}[(3m_r+2n_r)(m_r-2n_r)+8c_rp_r]\\
	a_2=\frac{2}{p_r^2}[p_r+m_ru],~~a_3=\frac{1}{p_r^2}[u^2+2kp_r].	
\end{gather*}
Eqn.(\ref{6}) reveals that the nonlinear chirp parameter $\alpha$ depends
on the model coefficients of the group velocity dispersion, SS and
the delayed nonlinear gain while $\beta$ depends on the group velocity 
dispersion and the velocity $u$ of the nonlinear mode. It is easy to check, 
with the help of the parametric conditions for the existence of 
all the solutions presented 
below, that the velocity depends on SS, delayed nonlinear gain and
the quintic Kerr effect. So the chirping can be controlled by changing these model coefficients. The feasibility of the proposed solutions in real applications depend on fixing the model parameters $q_i$, $c_i$, $n_i$ obeying Eq.(\ref{6}). In a ring fiber laser mode locked through nonlinear polarization rotation \cite{koma}, the coefficient $q_i$ can be adjusted to any value by a choice of the combination of the angles of the phase plates. Accordingly, for a given value of the solution parameters $u$ and $\alpha$, the values of $m_i$, $c_i$ and $n_i$ can be theoretically adjusted by using the appropriate relations in Eqn.(\ref{6}).
Now Eqn.(9) can be mapped to $\phi^6$ field 
equation \cite{behera} which enables us to obtain exact chirped dissipative 
solitary wave solutions of Eqn.(1).\\  
\subsection{Chirped solitary waves}
For all the solutions to (\ref{8})
presented below, there are different parametric conditions on $a_1, a_2, 
a_3$ for their existence \cite{behera}. The minimum number of nonzero model parameters required for their existence are $p_r$, $c_r$ and $m_r$ (or $n_r$).\\
\noindent I. {\bf Bright:}
~~\\
Eqn. (\ref{8}) admits the solution \cite{behera}
\begin{equation}\label{11}
	\rho=\frac{A~\rm{sech}(B\xi+\xi_0)}{\sqrt{1-D~\rm{tanh}^2(B\xi+\xi_0)}}\,.
\end{equation}
provided		
\begin{equation}\label{12}
		D< 1\,,~~B^2=-a_3,~a_2A^2=2(1+D)B^2,~~\frac{3a_2^2}{4a_3a_1}
		=\frac{(1+D)^2}{D}\,.
\end{equation}
We consider the solution when
$0<D<1$ and in that case $a_1, a_3 < 0, a_2 > 0$.
It is worth pointing out that in view of the identity
$\frac{\sinh(2\Delta)}{\sinh^2(\Delta)+\cosh^2(x)} = \tanh(x+\Delta) -
\tanh(x-\Delta)$, 
the solution (\ref{11}) can be re-expressed as a superposition of a kink
and an anti-kink solution, i.e.
\begin{equation}\label{sup}
	\rho = \frac{\sqrt{2\coth(2\Delta)} B}{\sqrt{a_2}}[\tanh(P+\Delta)
	-\tanh(P-\Delta)]^{1/2}\,,
\end{equation}
where $D = \tanh^2(\Delta), P = B\xi+\xi_0$. The dependence of the velocity, amplitude and chirp profile on different model parameters are shown in Fig.1. Figs.1(a) and (b) shows respectively, that the amplitude and chirping increases (decreases) as $m_r$ ($n_r$) increases but the velocity decreases (increases) as $m_r$ ($n_r$) increases. 
For increasing $c_r$, the chirping increases but the velocity and amplitude decreases (Fig.1(c)). Fig.1(b) and Fig.1(c) demonstrate respectively the switching from positive (negative) chirping to negative (positive) chirping as the $n_r$ ($c_r$) increases.\\
\begin{figure}[]
	\centering
	\subfloat[\label{}]{\includegraphics[width=5cm,height=4cm]{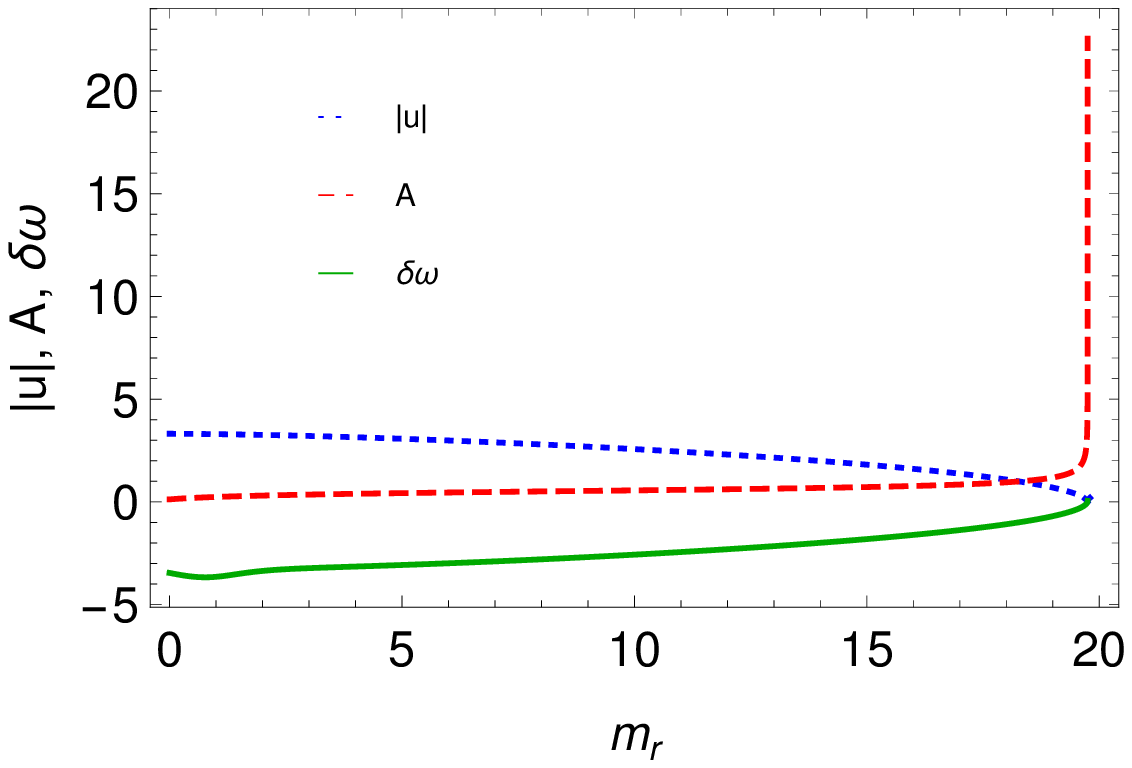}}
	~~
	\subfloat[\label{}]{\includegraphics[width=5cm,height=4cm]{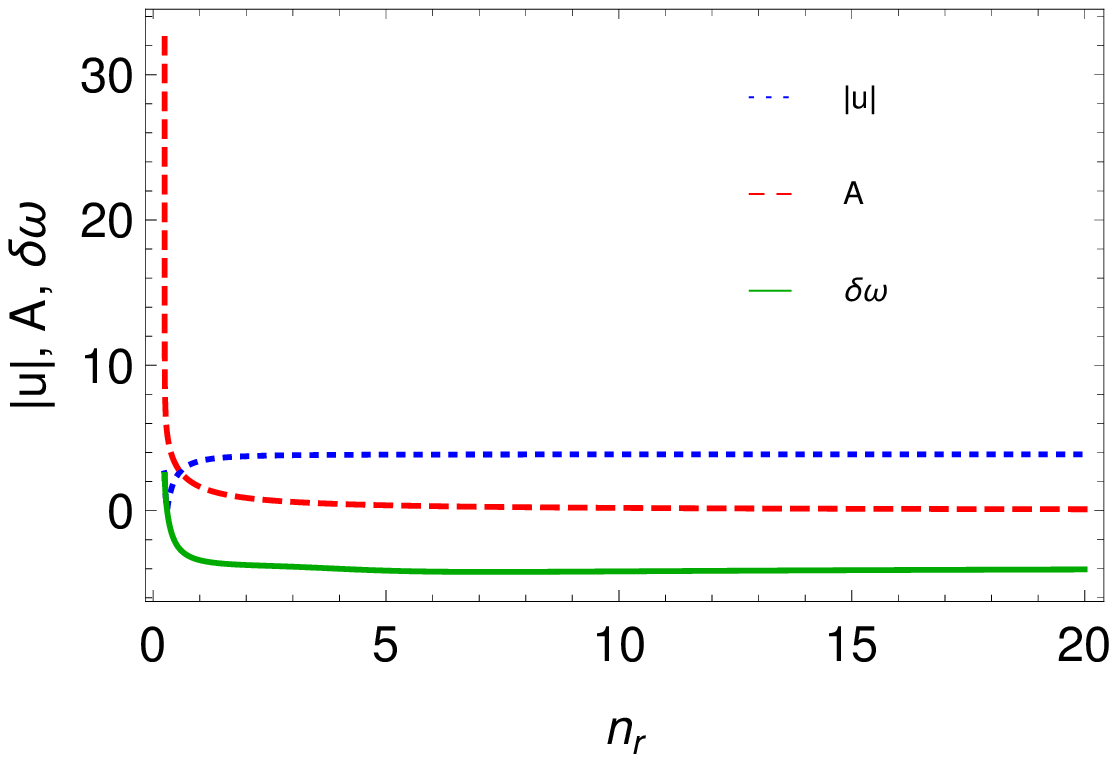}}
	~~
	\subfloat[\label{}]{\includegraphics[width=5cm,height=4cm]{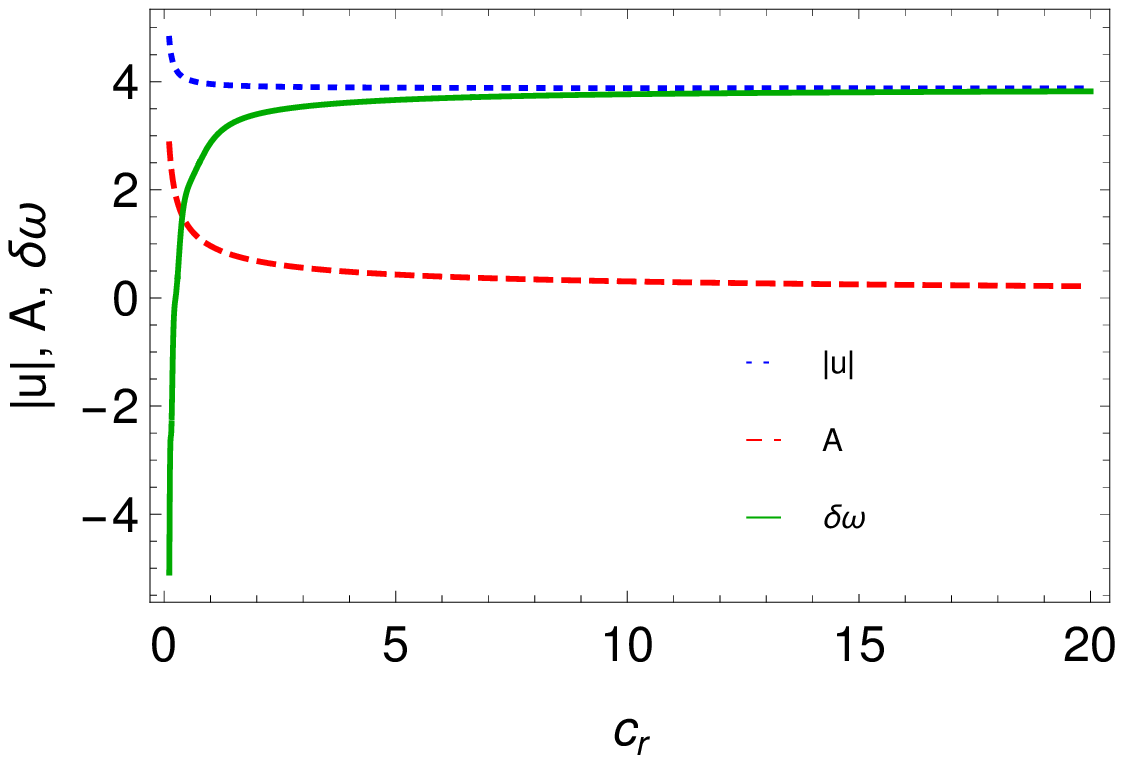}}\\
	
	~~	
	~~
	
	\subfloat[\label{}]{\includegraphics[width=5cm,height=4cm]{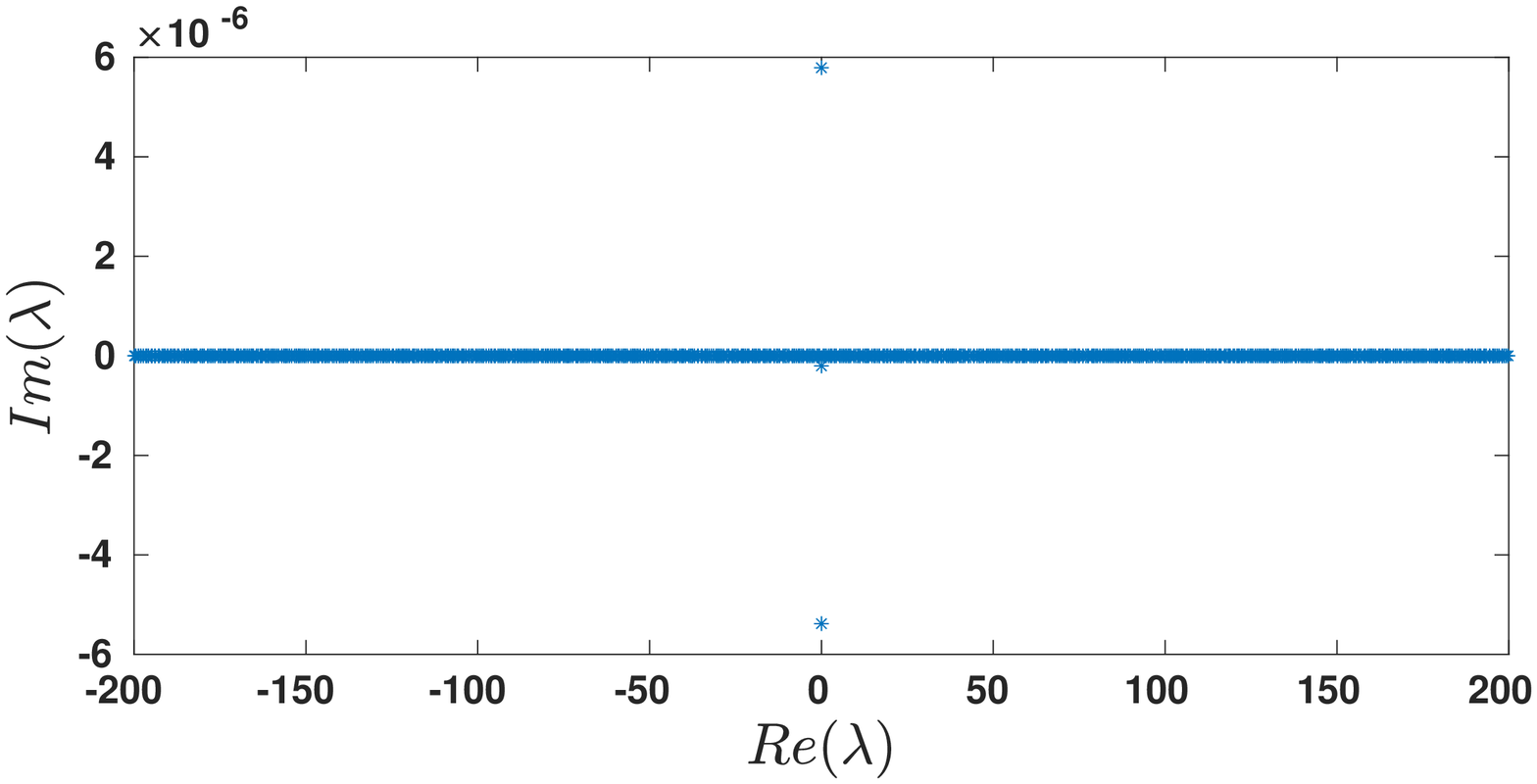}}
	~~
	\subfloat[\label{}]{\includegraphics[width=5cm,height=4cm]{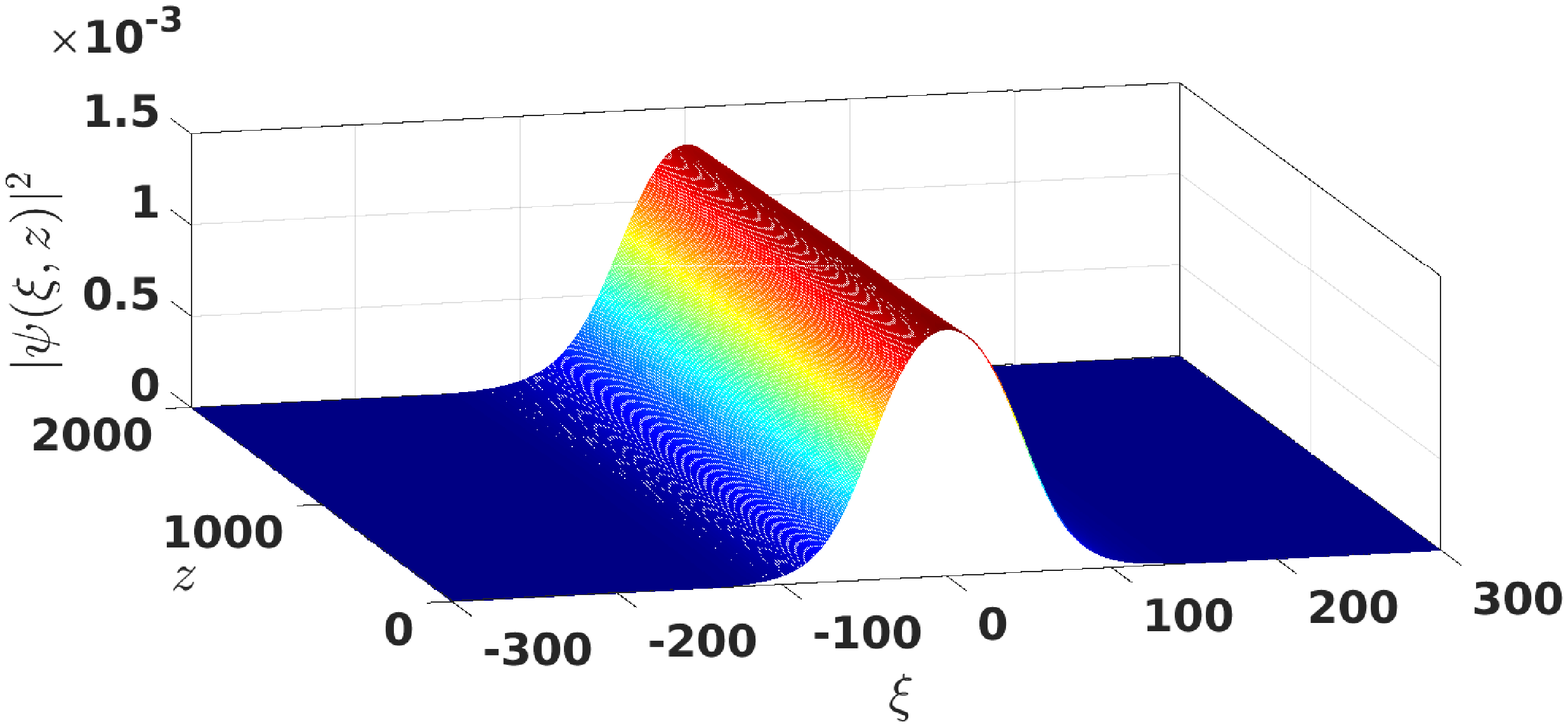}}
	\caption{For the solution(\ref{11}) with condition (\ref{12}) and $p_r=1$ (a), (b), (c) are plots of $|u|$ (blue dotted line), $A$ (red dashed line), $\delta\omega$ (green solid line) at $\xi=5,~\xi_0=0$ vs $m_r,~n_r,~c_r$ respectively for the parameter values (a) $n_r=9.9,~c_r=0.5,~D=0.9,~\kappa=-5.5$, (b) $m_r=0.4,~c_r=0.02,~D=0.9,~\kappa=-7.5$, (c) $m_r=0.6,n_r=0.2,~D=0.9,~\kappa=-7.5$. (d) Plot of eigen frequency and (e) propagation of nonlinear mode for $m_r=0.05\,,n_r=0.015,n_i=0.015,~D=0.8,~u=2.0,~\kappa=-2.0005$.}
\end{figure}
~~\\
II. {\bf Dark:}\\
\begin{equation}\label{16}
	\rho=\frac{A~\rm{tanh}(B\xi+\xi_0)}{\sqrt{1-D~\rm{tanh}^2(B\xi+\xi_0)}}
\end{equation}
is an exact solution to eqn.(\ref{8}) provided
\begin{equation}\label{17}
	\begin{split}
		D<1\,,~~(3D-2)B^2=-a_3,\\a_2A^2=2(3D-1)(1-D)B^2,~
		\frac{3a_2^2}{4a_3a_1}=\frac{(3D-1)^2}{D(3D-2)}.
	\end{split}	
\end{equation}
We choose $0<D<\frac{1}{3}$ and in that case $a_1, a_2<0,~~a_3>0$.

The amplitude and chirping initially decrease as $m_r$ increases and start increasing when a particular value of $m_r$ is reached while the velocity initially decreases as $m_r$ increases but saturates for higher values of $m_r$ as shown in (Fig.2(a)). When $n_r$ increases, the velocity and amplitude decrease, while the chirping increases at first, but then decreases as $n_r$ approaches a certain value (Fig.2(b)). The velocity, amplitude and chirping, all  decrease as $c_r$ increases (Fig.2(c)).\\

\begin{figure}[]
	\centering
	\subfloat[\label{}]{\includegraphics[width=5cm,height=4cm]{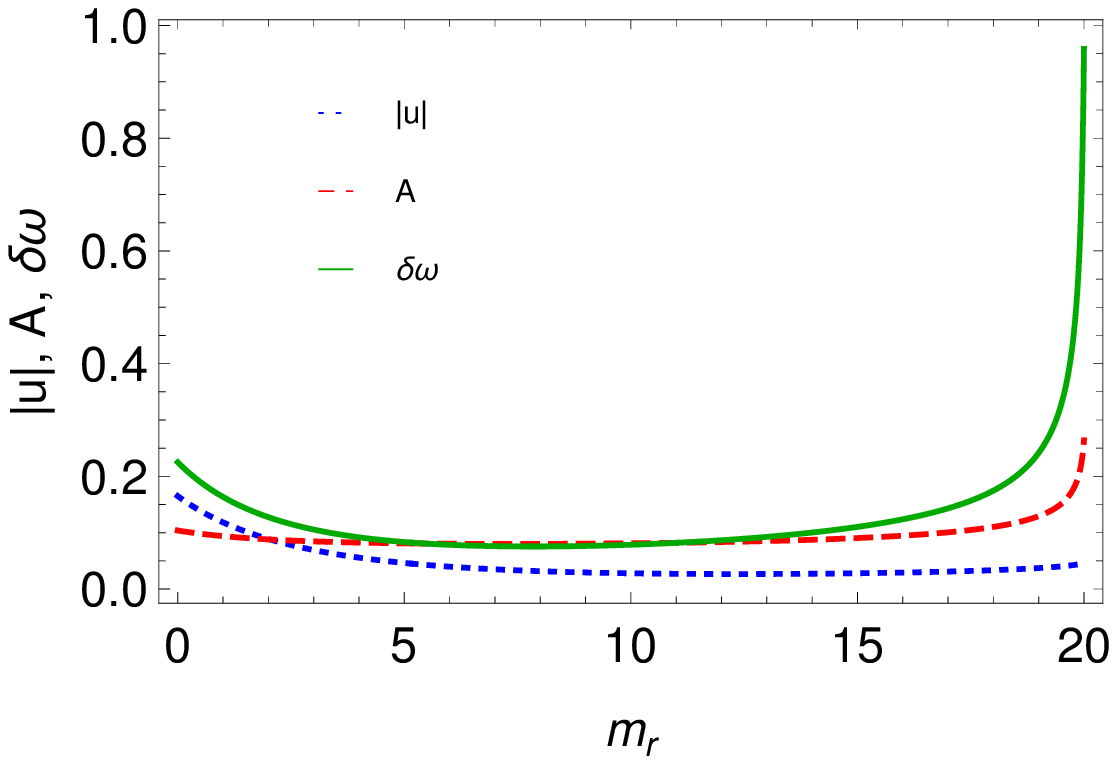}}
	~~
	\subfloat[\label{}]{\includegraphics[width=5cm,height=4cm]{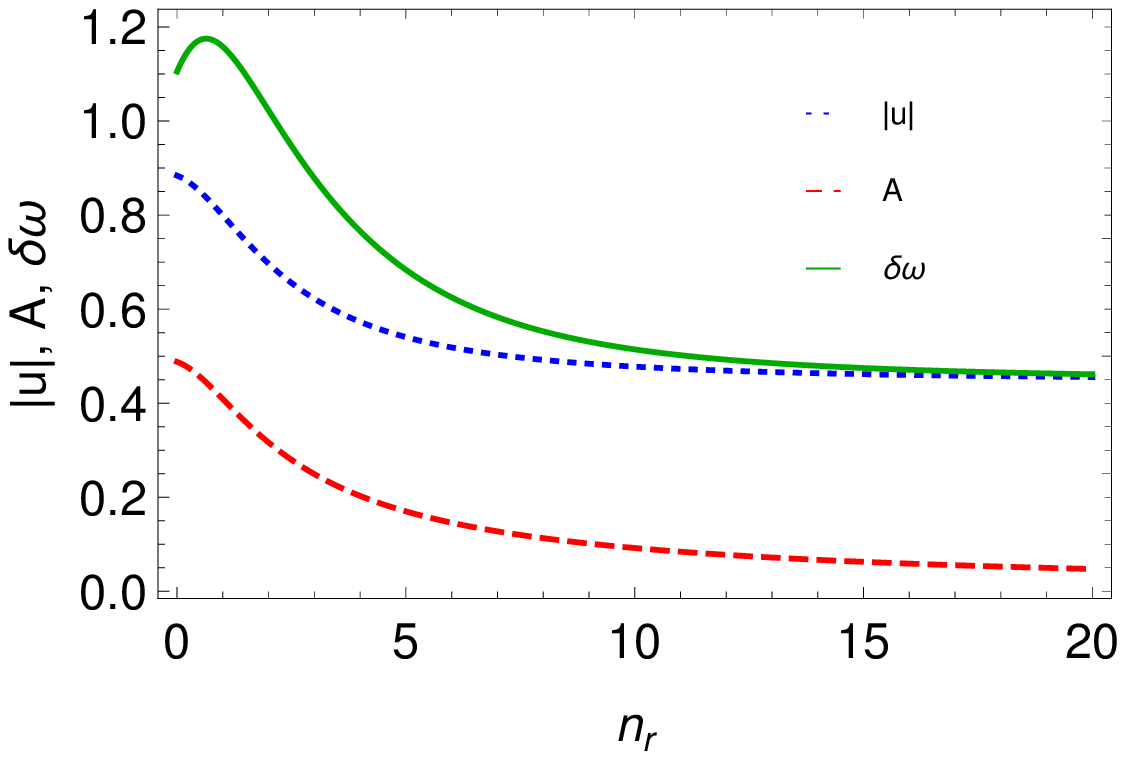}}
	~~
	\subfloat[\label{}]{\includegraphics[width=5cm,height=4cm]{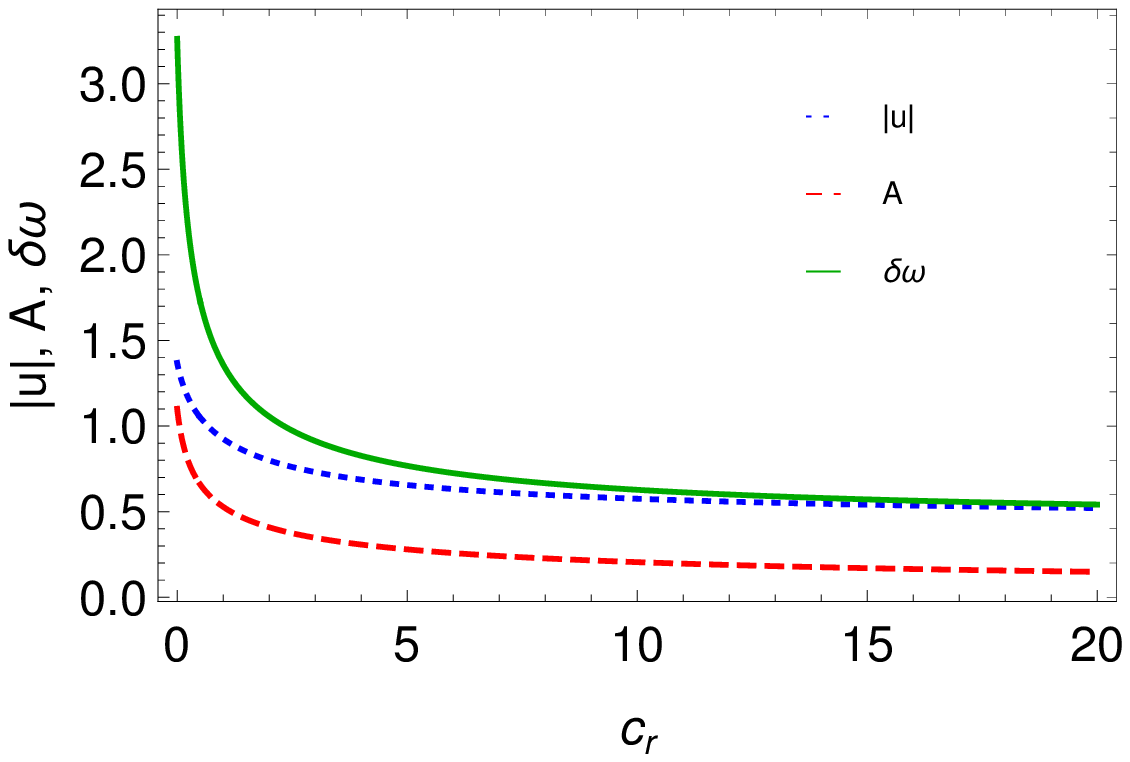}}\\
	~~
	\subfloat[\label{}]{\includegraphics[width=5cm,height=4cm]{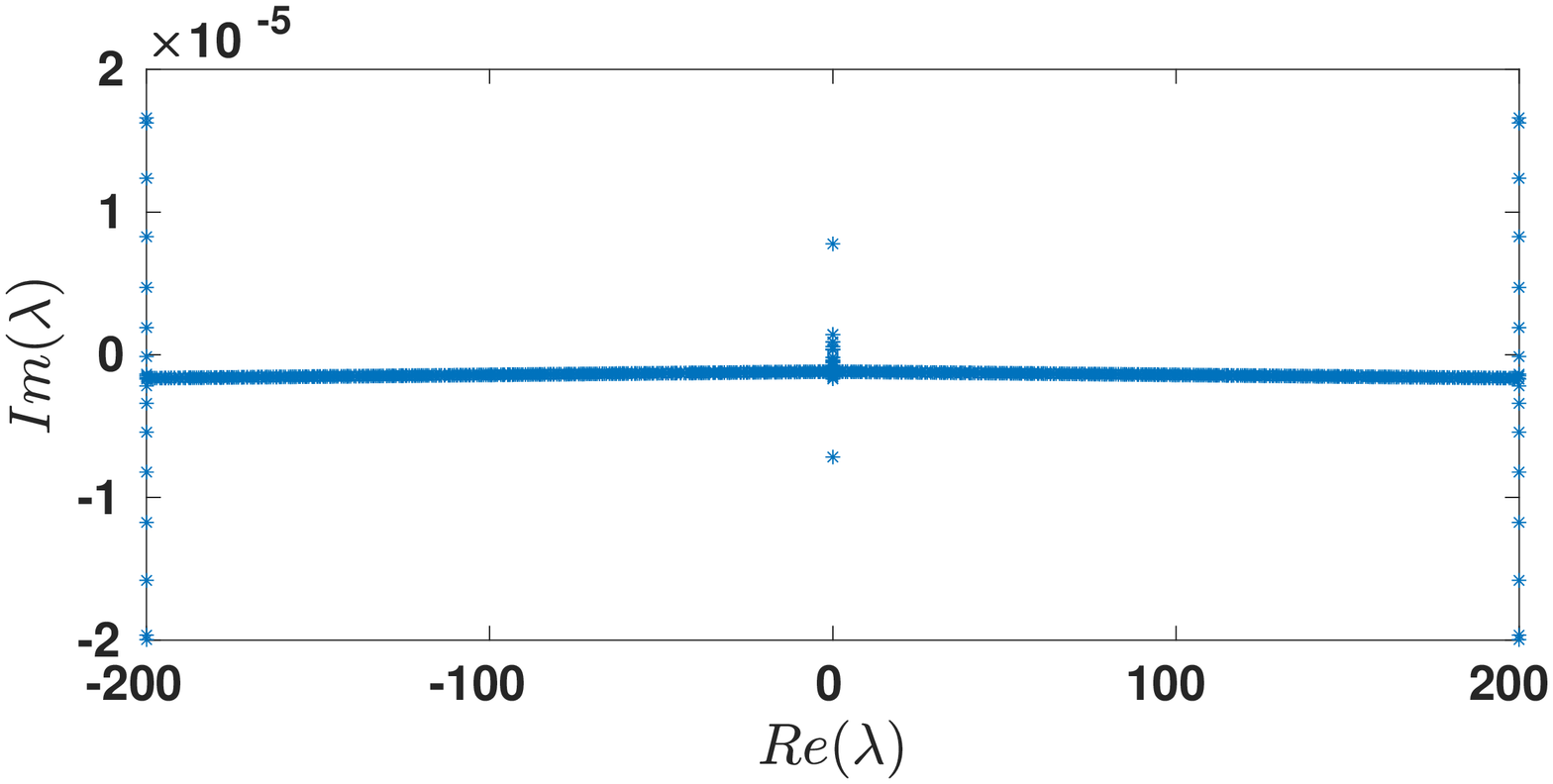}}
	~~
	\subfloat[\label{}]{\includegraphics[width=5cm,height=4cm]{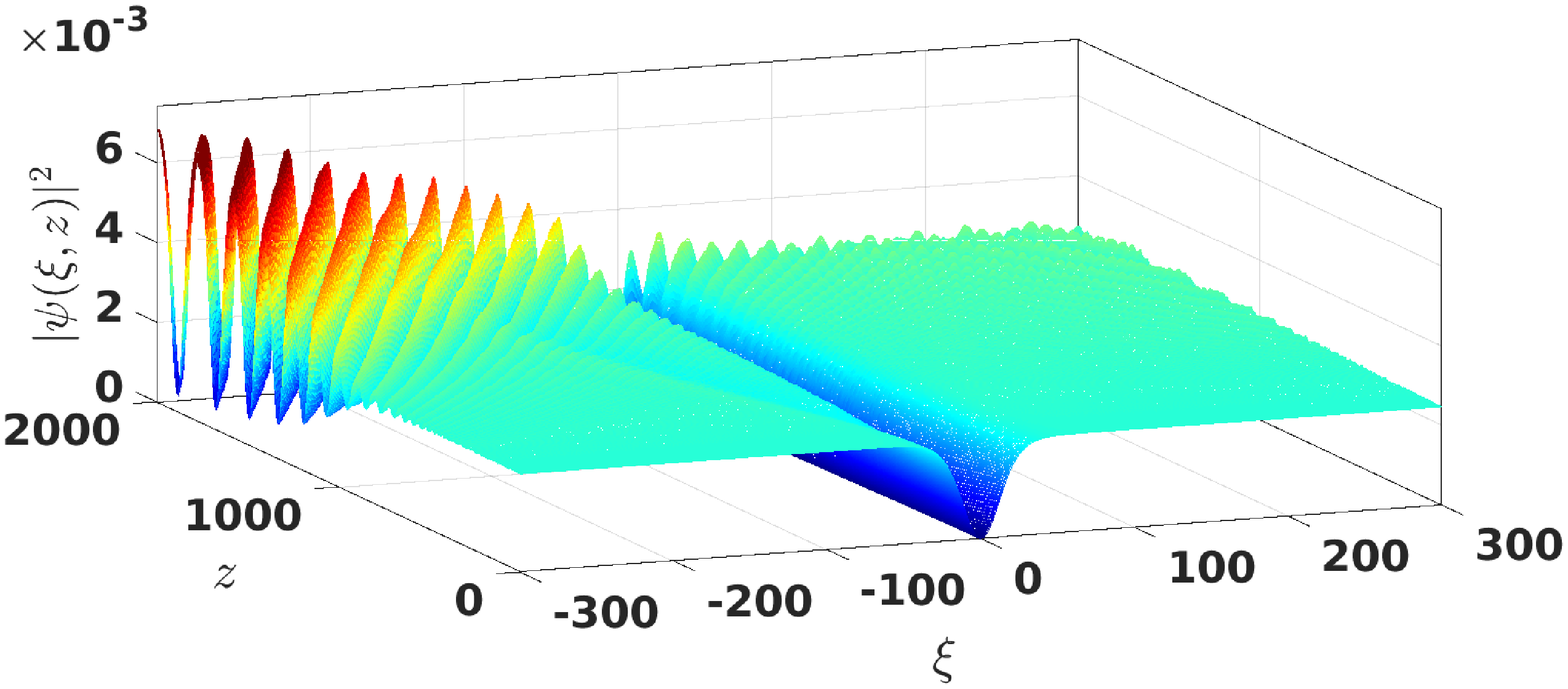}}
	\caption{For the solution(\ref{16}) with condition (\ref{17}) and $p_r=-1,~D=0.2$ , (a), (b), (c) are plots of $|u|$ (blue dotted line), $A$ (red dashed line), $\delta\omega$ (green solid line) at $\xi=5,~\xi_0=0$ vs $m_r,~n_r,~c_r$ respectively for the parameter values (a) $n_r=9.5,~c_r=10.5,~\kappa=-0.01$, (b) $m_r=0.5,~c_r=1,\kappa=0.1$, (c) $m_r=0.5,n_r=0.5,\kappa=0.1$. (d) Plot of eigen frequency and (e) propagation of nonlinear mode for $m_r=2.5,~n_r=0.05,~n_i=0.1,~u=0.12,~\kappa=0.0042$.}
\end{figure}
~~\\
III. {\bf Kink and antikink solitary waves}\\
(a).{\bf Kink}:

\begin{equation}\label{14}
	\rho=A\sqrt{1+\rm{tanh}(B\xi+\xi_0)}
\end{equation}
is an exact solution to eqn.(\ref{8}) if the following constraints are satisfied:
\begin{equation}\label{15}
	B^2=-a_3,~~~~~~~~~~~A^2=-\frac{2a_3}{a_2},~~~~~~~~~~a_2^2=\frac{16}{3}a_3a_1
\end{equation}
(b). {\bf Antikink}:
\begin{equation}\label{14a}
	\rho=A\sqrt{1-\rm{tanh}(B\xi+\xi_0)}
\end{equation}
is an exact solution to eqn.(\ref{8}) provided the constraints given by the eqn.(\ref{15}) are satisfied. \\
The amplitude and chirp profiles of solutions (\ref{14}) and (\ref{14a}) for different model parameters are shown in Fig.3 and Fig.4 respectively.
The velocity decreases as $m_r$ increases (Figs.3(a)) but the velocity increases for increasing values of $n_r$ and $c_r$ (Figs.3(b), (c)). Fig.3(a) shows that, as $m_r$ increases, the chirping increases while the amplitude initially increases slightly and then saturates for higher values of $m_r$. For increasing $n_r$ (Fig.3(b)) ($c_r$, Fig.3(c)), the amplitude and chirping, both decrease as $n_r$ ($c_r$) increases but saturate for higher values of $n_r$ ($c_r$). 
\begin{figure}[]
	\centering
	\subfloat[\label{}]{\includegraphics[width=5cm,height=4cm]{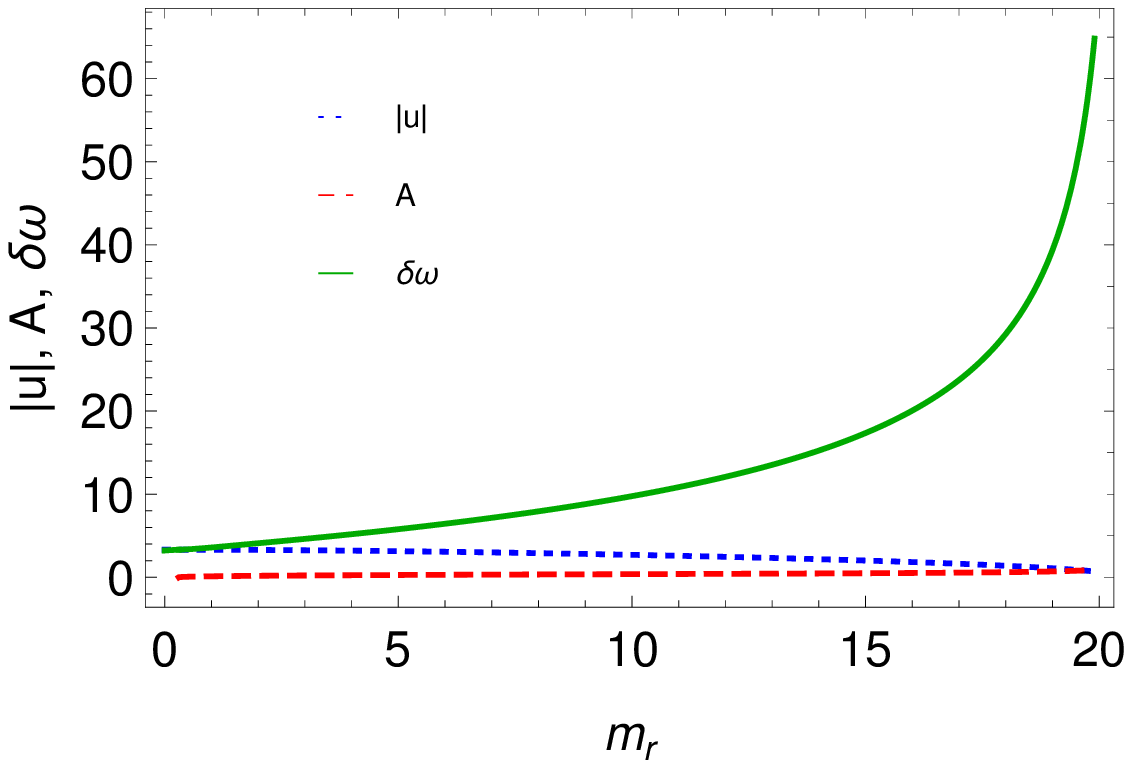}}
	~~
	\subfloat[\label{}]{\includegraphics[width=5cm,height=4cm]{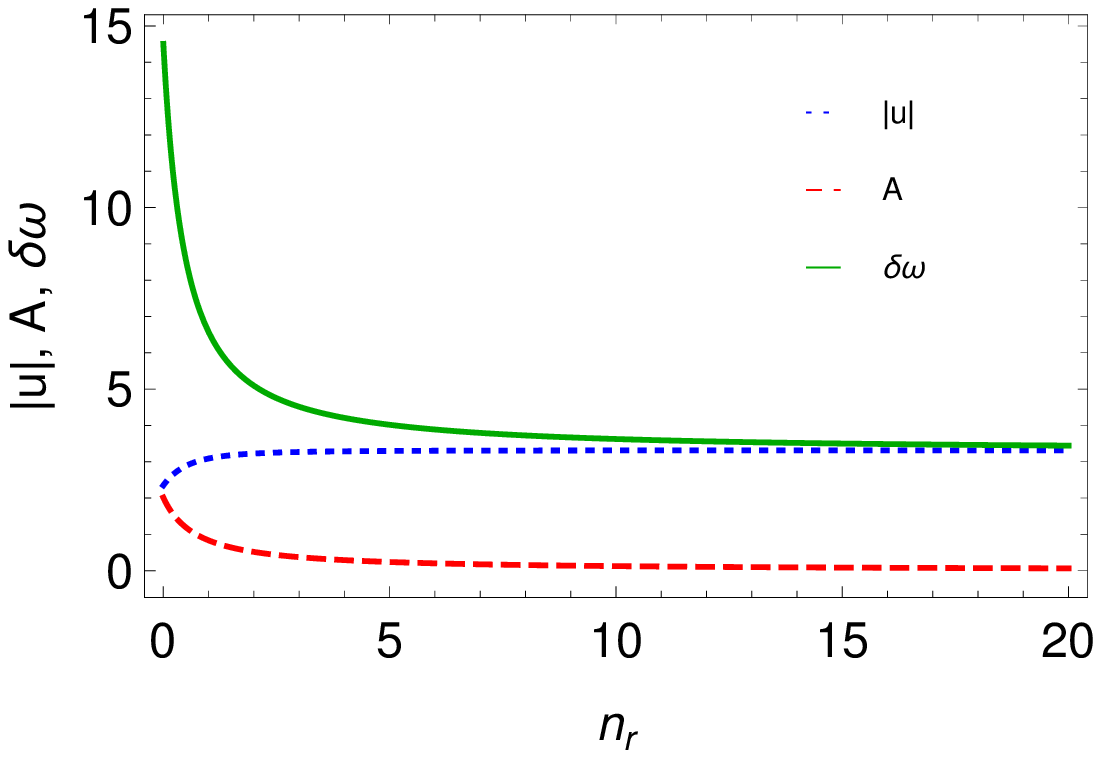}}
	~~
	\subfloat[\label{}]{\includegraphics[width=5cm,height=4cm]{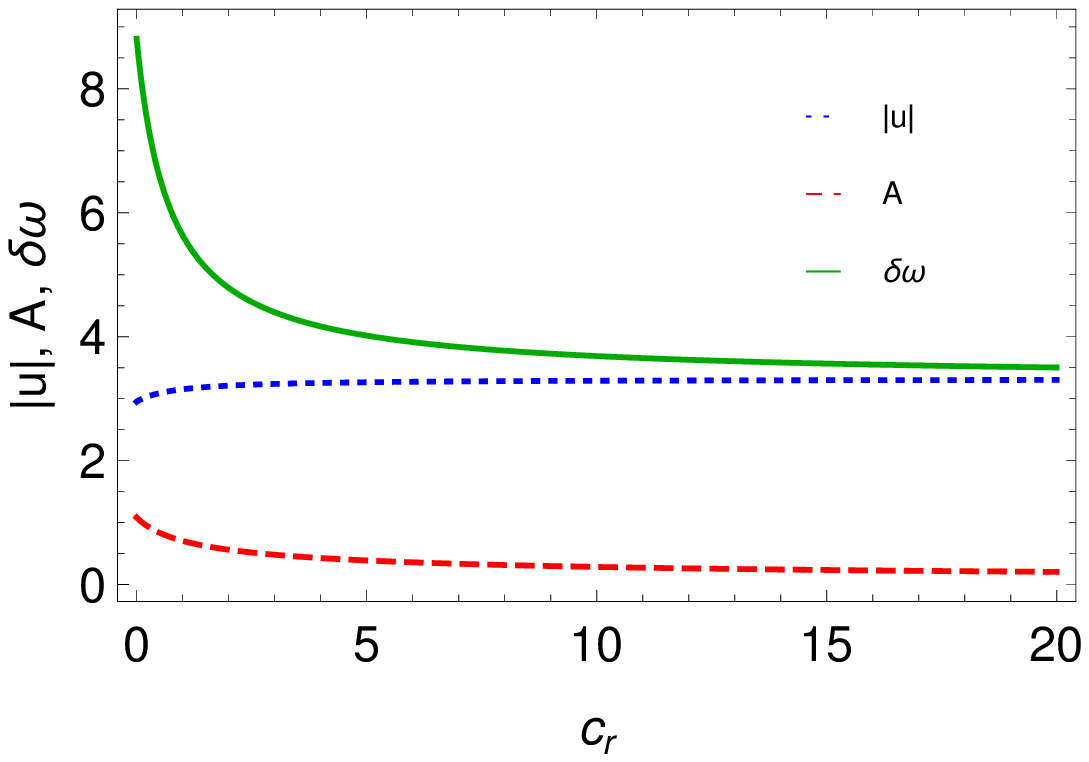}}\\
	
	~~
	\subfloat[\label{}]{\includegraphics[width=5.0cm,height=4cm]{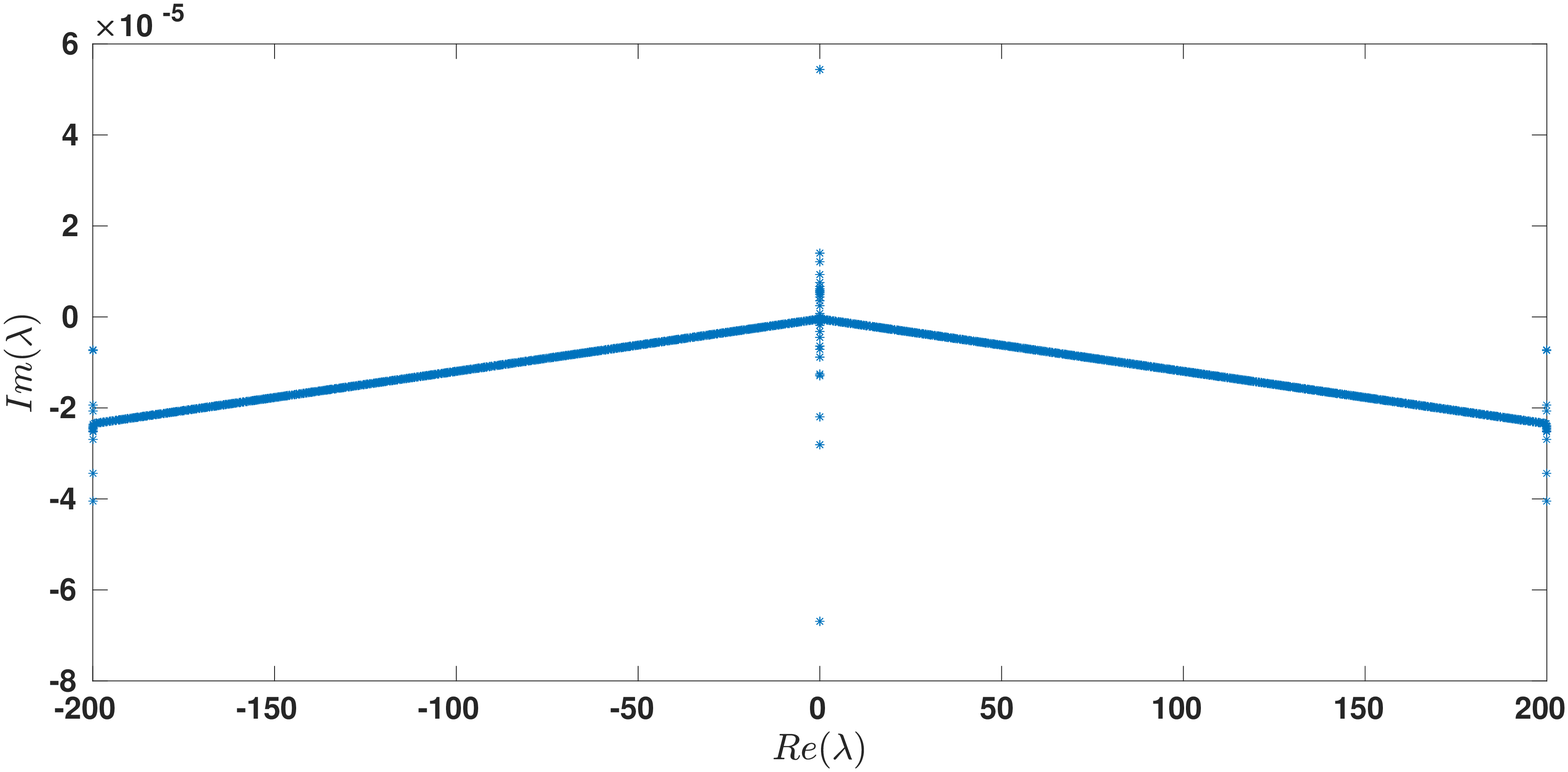}}
	~~
	\subfloat[\label{}]{\includegraphics[width=5.0cm,height=4cm]{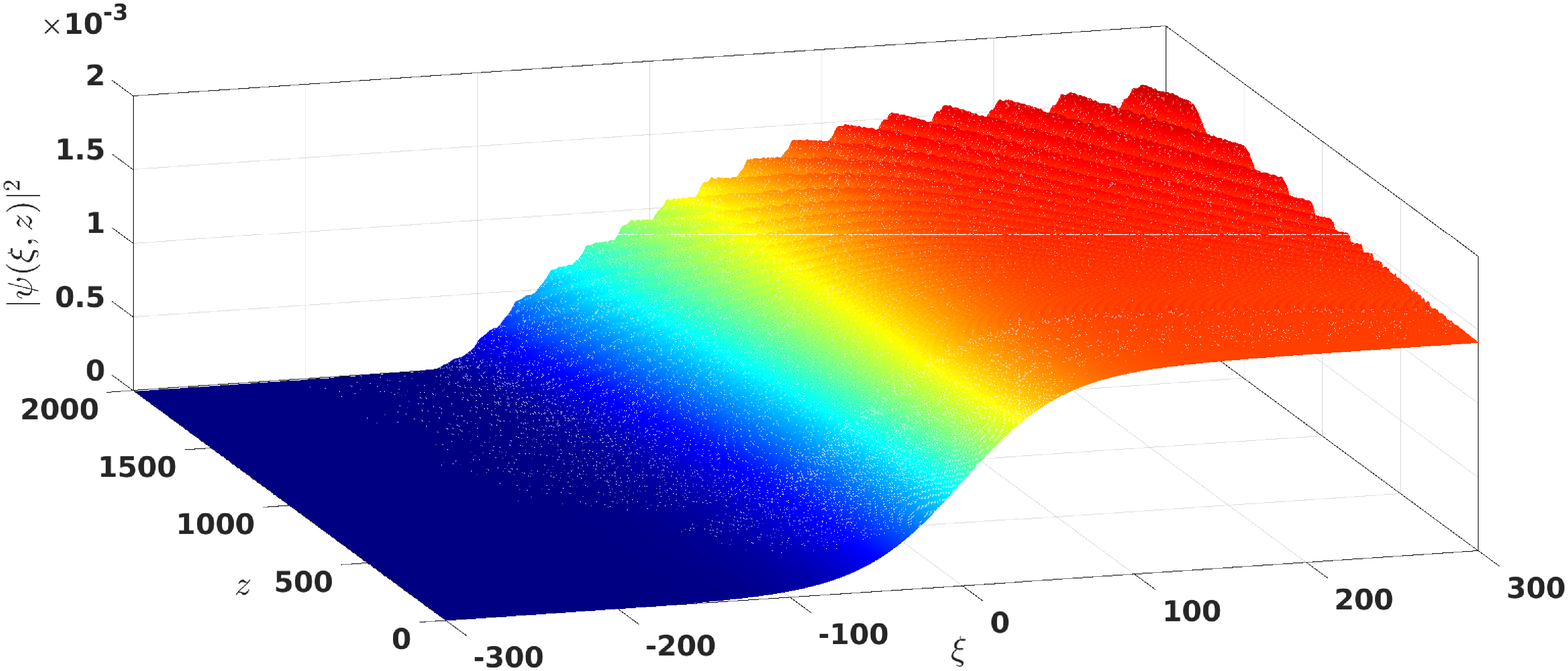}}\\
	
	\caption{For the solution(\ref{14}) with conditions (\ref{15}) with $p_r=-1$  (a), (b), (c) are plots of $|u|$ (blue dotted line), $A$ (red dashed line), $\delta\omega$ (green solid line) at $\xi=5,~\xi_0=0$ vs $m_r,~n_r,~c_r$ respectively for $\kappa=5.5$ and the other parameters are (a) $n_r=9.8,~c_r=9.0$, (b) $m_r=1.0,~c_r=0.5$, (c) $m_r=1.0,n_r=1.0$. (d) Plot of eigen frequency and (d) propagation of nonlinear mode for $m_r=11.26\,,n_r=-9.01,n_i=0.1,~u=0.1199,~\kappa=0.0073$.
	}
\end{figure}
When $m_r$ increases, the chirping initially increases, but starts decreasing as $m_r$ reaches a certain value (Fig.4(a)). The chirping increases as $n_r$ and $c_r$ increase, as demonstrated in Figs. 4(b) and 4(c) respectively. As qualitative behaviours of the speed and amplitude of the antikink solitary wave (\ref{14a}) with respect to $m_r, n_r, c_r$ are the same as those of the kink solution (\ref{14}) we refrain ourselves from giving the plots. 

\begin{figure}[]
	\centering
	\subfloat[\label{}]{\includegraphics[width=5cm,height=4cm]{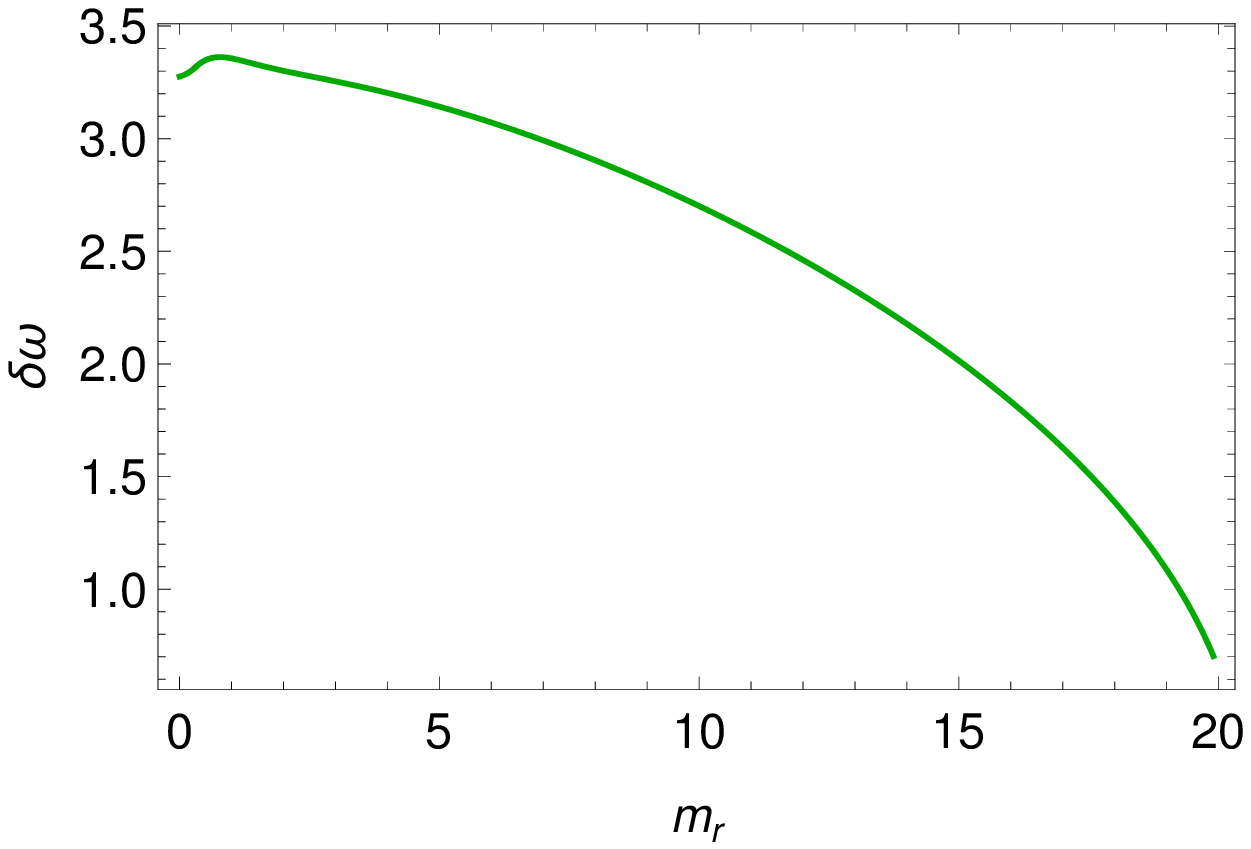}}
	~~
	\subfloat[\label{}]{\includegraphics[width=5cm,height=4cm]{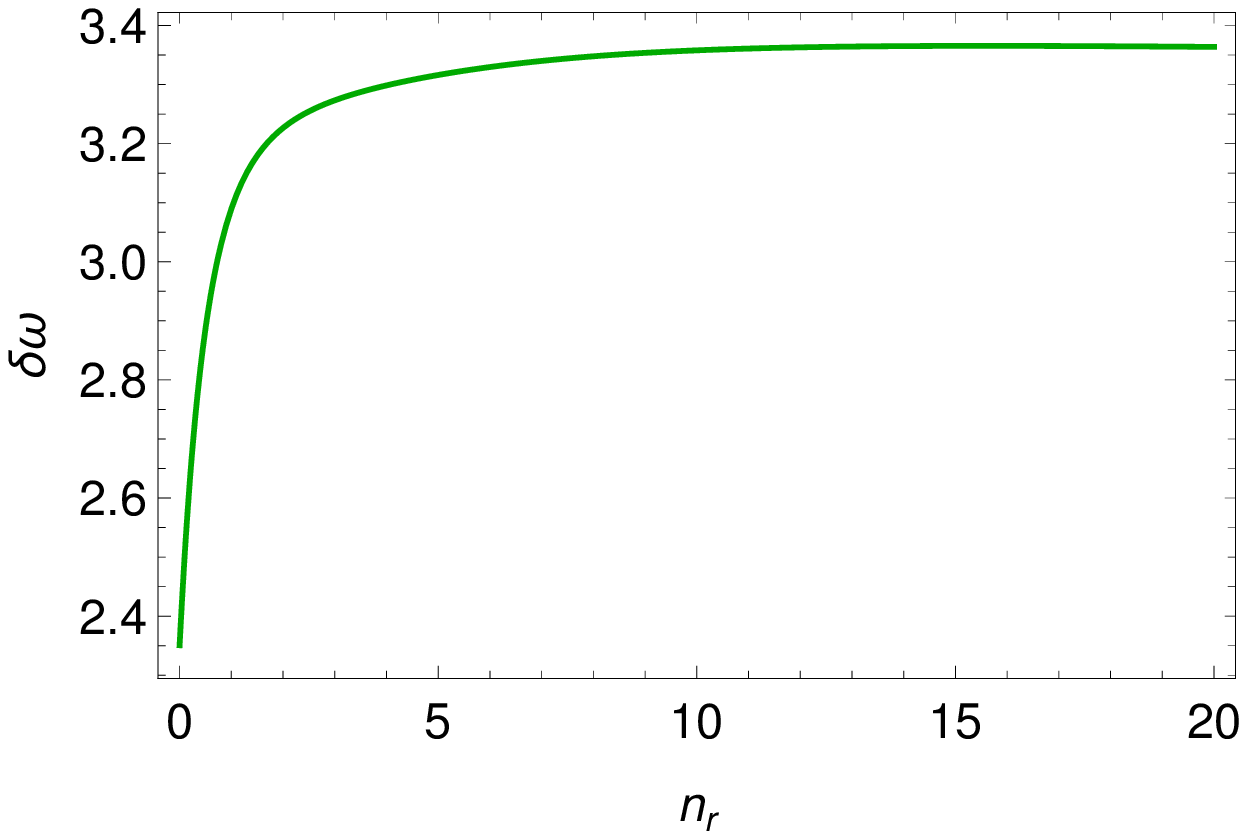}}
	~~
	\subfloat[\label{}]{\includegraphics[width=5cm,height=4cm]{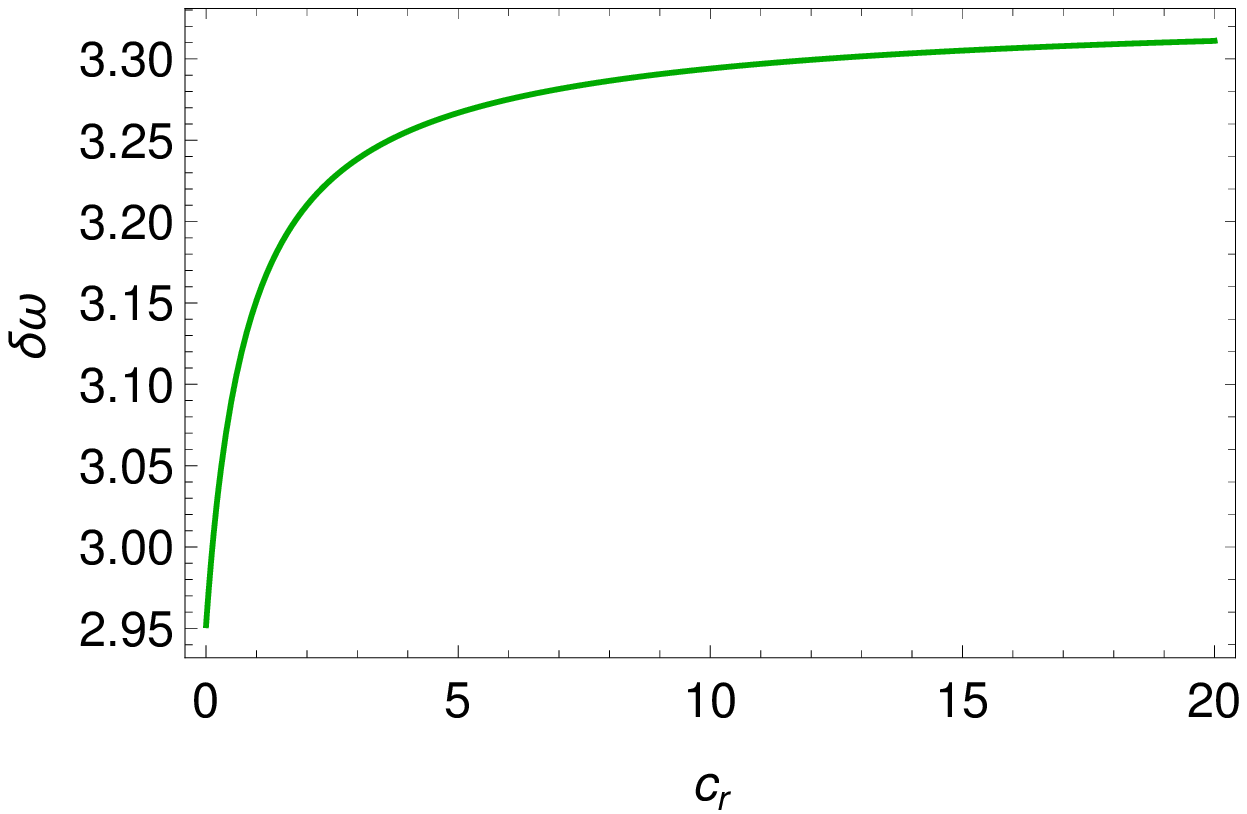}}\\
	~~
	~~
	\subfloat[\label{}]{\includegraphics[width=5cm,height=4cm]{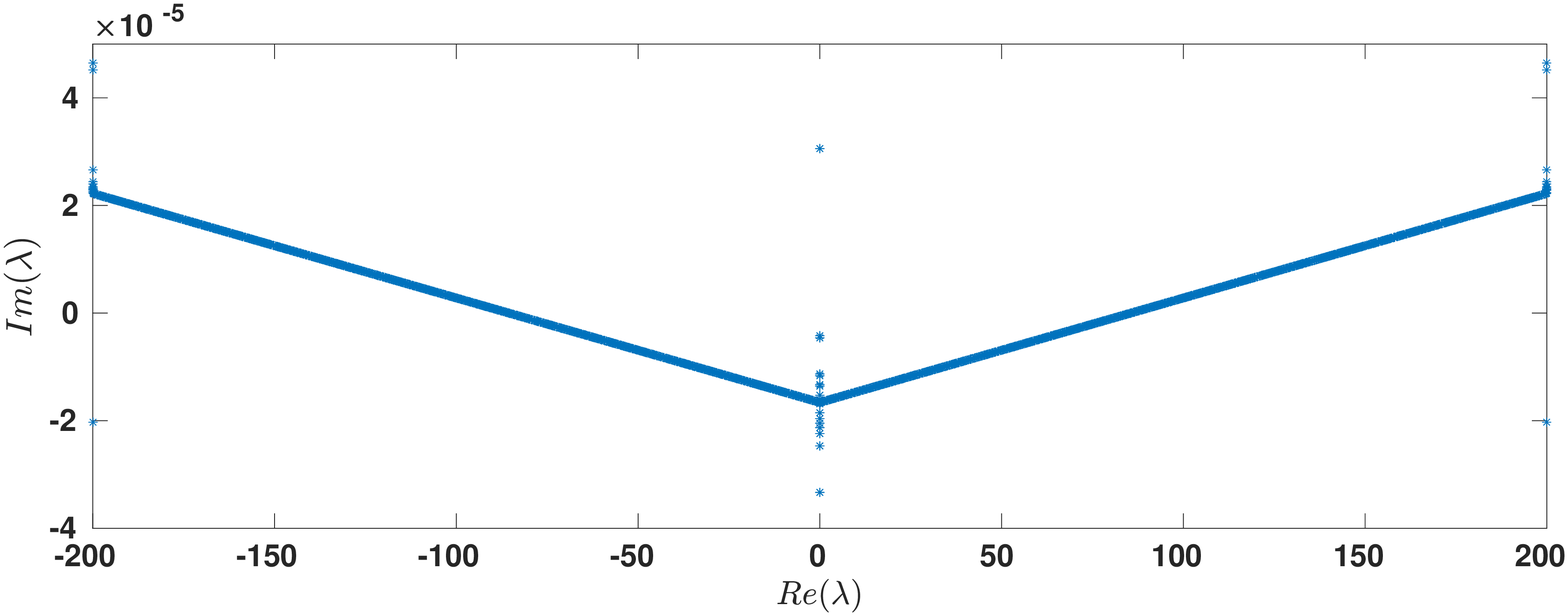}}
	~~
	\subfloat[\label{}]{\includegraphics[width=5cm,height=4cm]{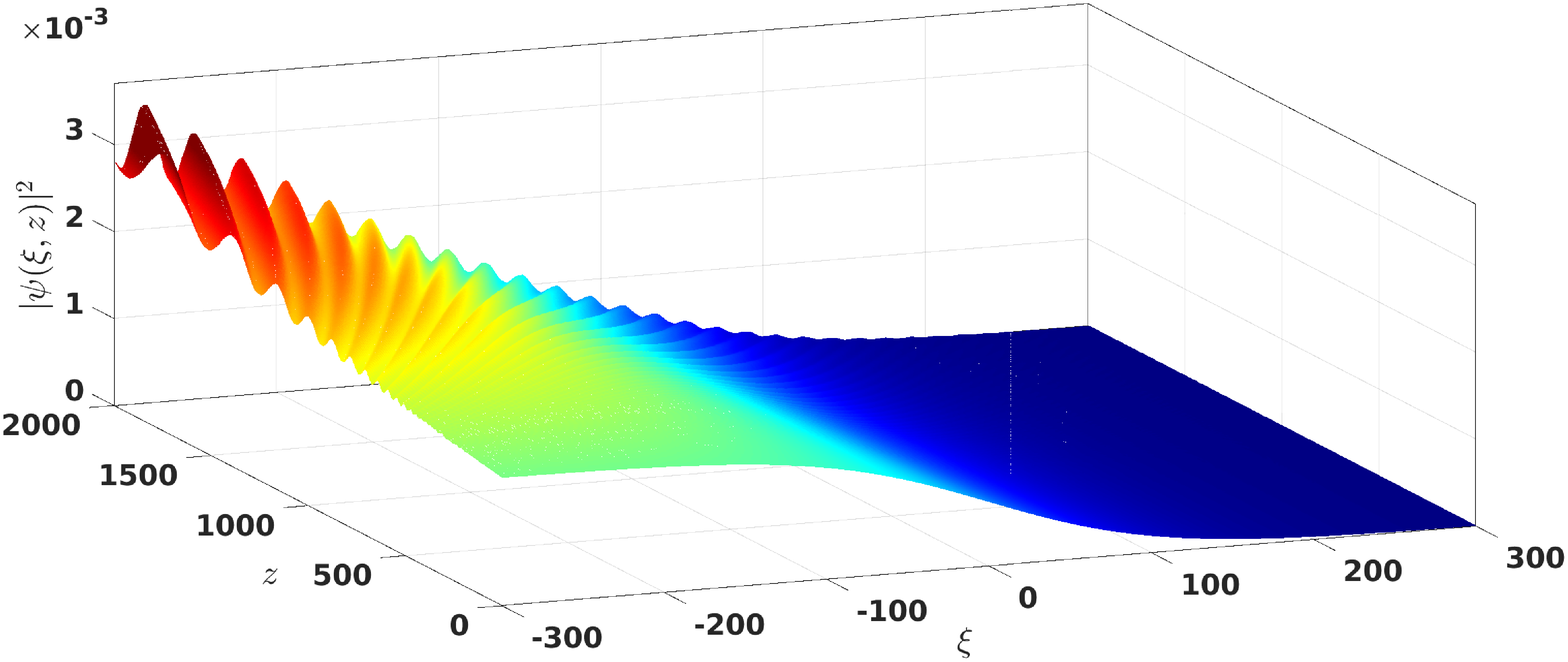}}\\
	
	\caption{For the solution(\ref{14a}) with conditions (\ref{15}) taking $p_r=-1$ (a), (b), (c) are plots of  $\delta\omega$ (green solid line) at $\xi=5,~\xi_0=0$ vs $m_r,~n_r,~c_r$ respectively for $\kappa=5.5$ and other parameters are (a) $n_r=9.8,~c_r=9.0$, (b) $m_r=1.0,~c_r=0.5$, (c) $m_r=1.0,n_r=1.0$. (d) plot of eigen frequency and (e) propagation of nonlinear mode for $m_r=15.0\,,n_r=-6.0,n_i=1.5,~u=0.0733,~\kappa=0.0027$.
	}
\end{figure}
~~\\
IV. {\bf Grey and Antidark solitary waves}\\
Yet another solution to eqn.(\ref{8}) is
\begin{equation}\label{19}
	\rho=\frac{A}{\sqrt{1-D~\rm{tanh}^2(B\xi+\xi_0)}}
\end{equation}
subject to the following constraints 
\begin{equation}\label{20}
	\begin{split}
		D < 1\,,~~(3-2D)B^2=-a_3D\\
		Da_2A^2=2(3-D)(1-D)B^2,~\frac{3a_2^2}{4a_3a_1}=\frac{(3-D)^2}{3-2D}\,.				
	\end{split}
\end{equation}
Depending on the value of $D$ one gets grey or antidark solitary waves.\\
(a) {\bf Grey}: For this, the parametric conditions are
\begin{equation}\label{20a}
	0<D<1,~~~~~~~a_1<0,~~~~~~~~a_2>0,~~~~~~~~~a_3<0
\end{equation}
(b). {\bf Antidark}: In this case the constraints are
\begin{equation}\label{20b}
	D<0,~~~~~~a_1>0,~~~~~~a_2<0,~~~~~a_3>0
\end{equation}	
For the grey solitary wave (\ref{20a}), the velocity decreases for increasing $m_r$ (Fig.5(a)) and $n_r$ (Fig.5(b))  but it increases as $c_r$ increases (Fig.5(c)). The amplitude decreases for increasing $m_r$ (Fig.5(a)), $n_r$ (Fig.5(b)), $c_r$ (Fig.5(c)). The chirping increases as $m_r$ increases (Fig.5(a)) but it decreases as $c_r$ increases (Fig.5(c)). As $n_r$ increases, the chirping initially decreases then after increasing slightly, it again decreases (Fig.5(b)).\\ 
Fig.6(a) shows that the velocity and chirping of the antidark solitary wave (\ref{20b}) increases as $m_r$ increases while the amplitude initially decreases but starts increasing when a certain value of $m_r$ is reached. The amplitude decreases as $n_r$ (Fig.6(b)), $c_r$ (Fig.6(c)) increases. The velocity slightly increases as $n_r$ and $c_r$ increase but it saturates for higher values of $n_r, c_r$(Figs.6(b), 6(c)). The chirping initially increases but subsequently decreases for increasing $n_r$ but it saturates for higher values of $n_r$ (Fig.6(b)) while for increasing $c_r$, chirping initially decreases and then it saturates (Fig.6(c)).

\begin{figure}[]
	\centering
	\subfloat[\label{}]{\includegraphics[width=5cm,height=4cm]{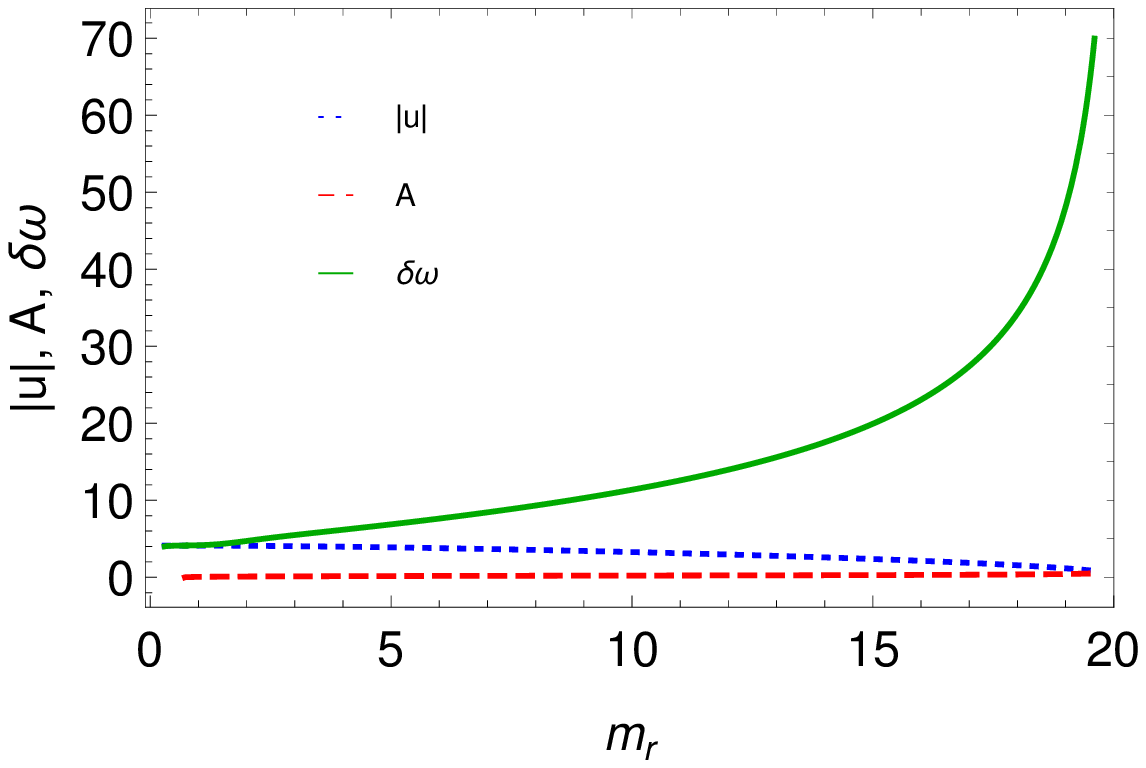}}
	~~
	\subfloat[\label{}]{\includegraphics[width=5cm,height=4cm]{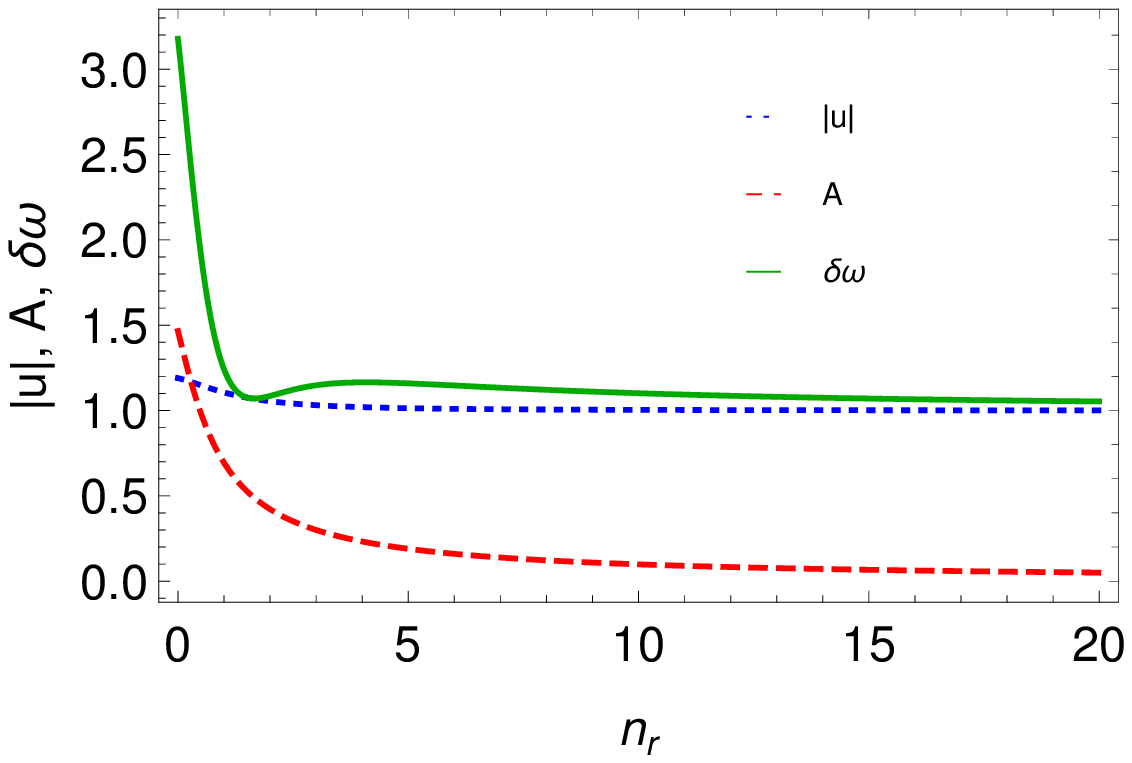}}
	~~
	\subfloat[\label{}]{\includegraphics[width=5cm,height=4cm]{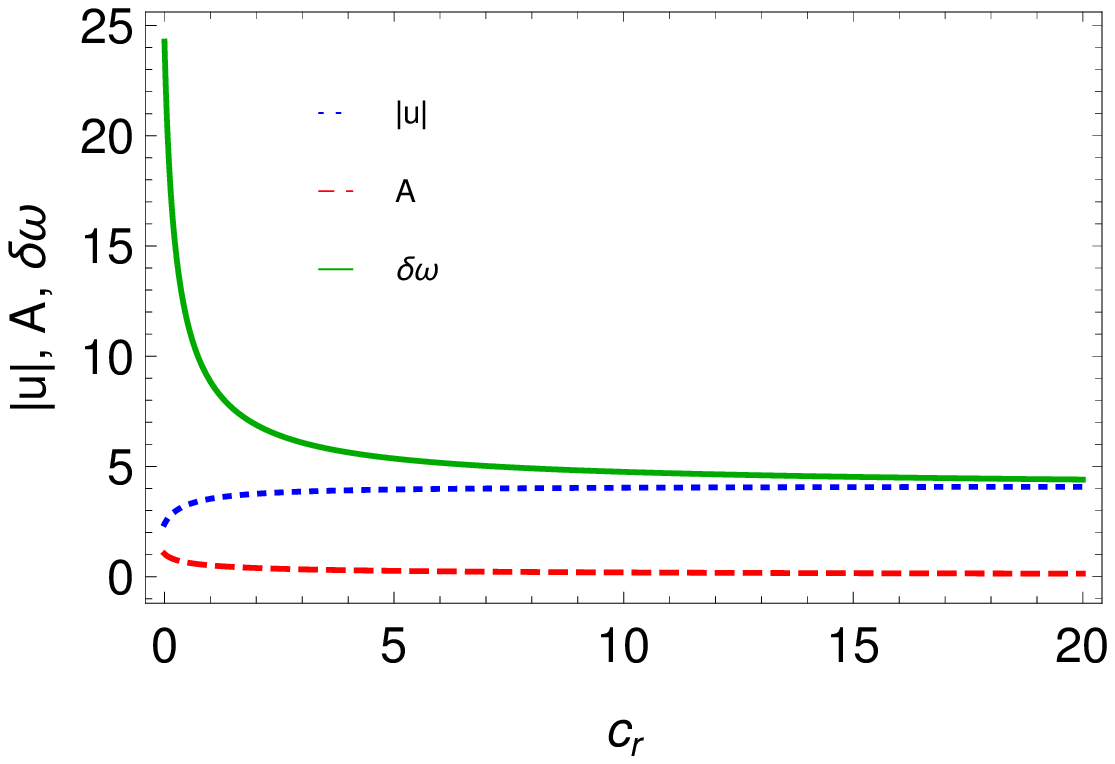}}\\
	
	~~
	~~
	\subfloat[\label{}]{\includegraphics[width=5.0cm,height=4cm]{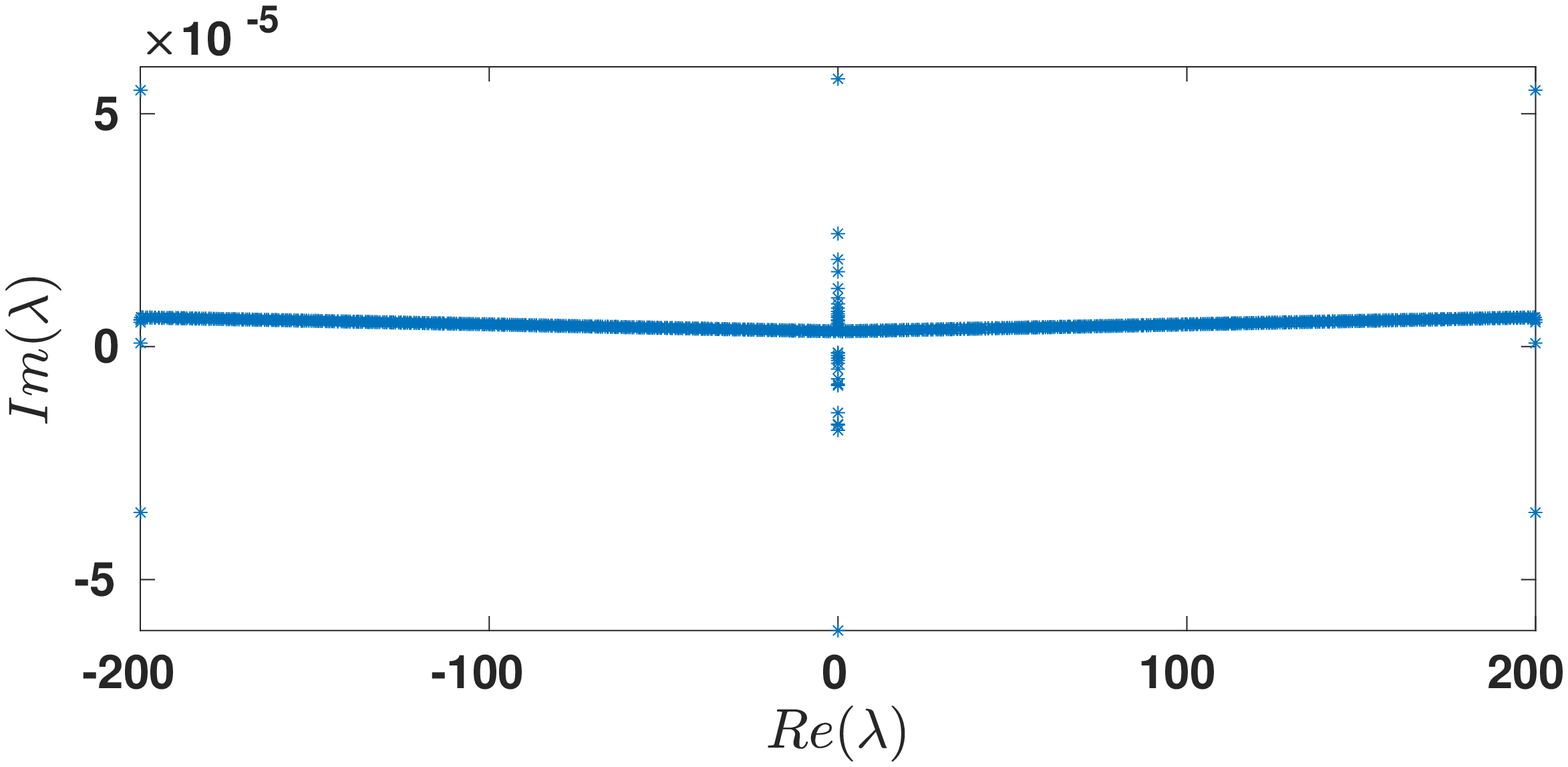}}
	~~
	\subfloat[\label{}]{\includegraphics[width=5.0cm,height=4cm]{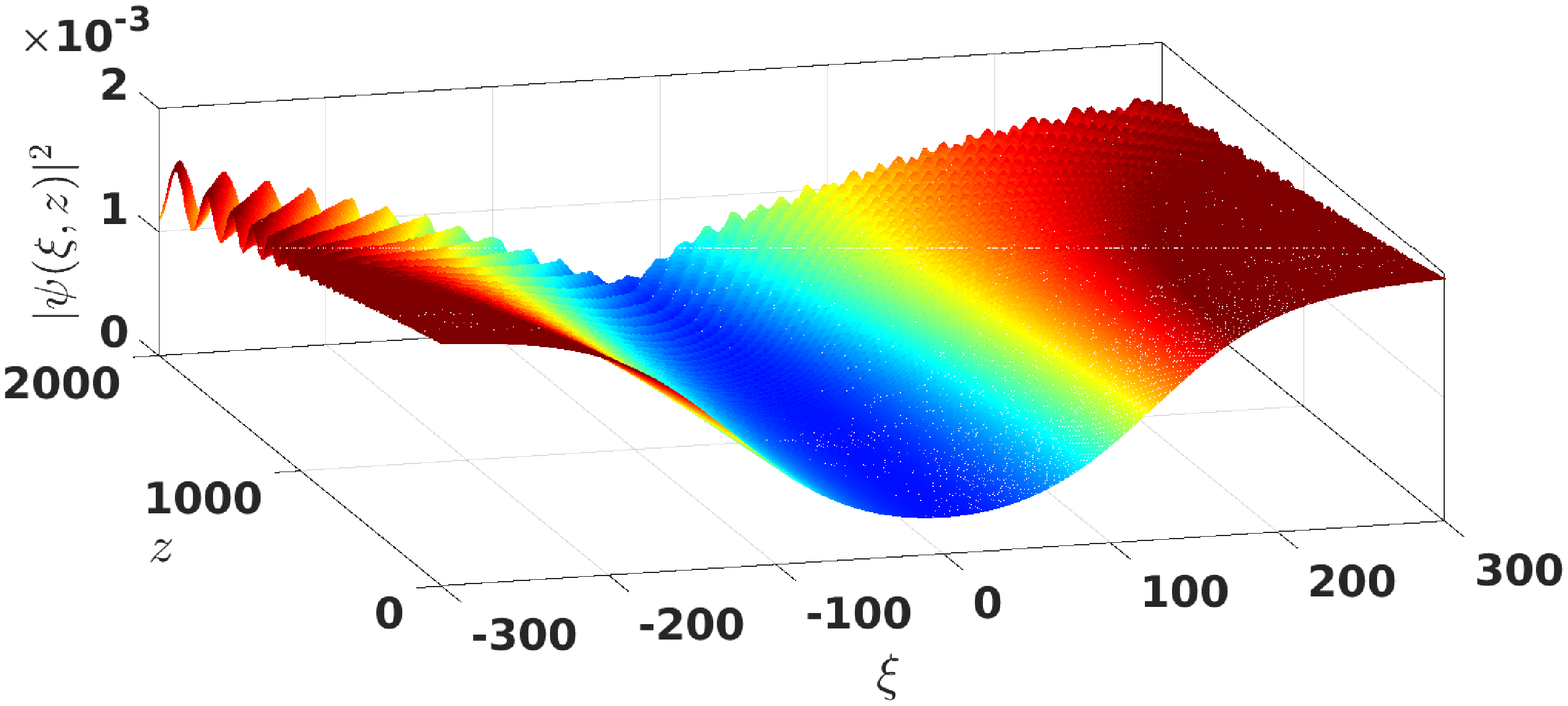}}\\
	
	\caption{For the solution(\ref{19}) with conditions (\ref{20}) and (\ref{20a}) with  $p_r=-1,~D=0.85$, (a), (b), (c) are plots of $|u|$ (blue dotted line), $A$ (red dashed line), $\delta\omega$ (green solid line) at $\xi=5,~\xi_0=0$ vs $m_r,~n_r,~c_r$ respectively for the parameters (a) $n_r=9.8,~c_r=5.5,~\kappa=8.5$, (b) $m_r=0.8,~c_r=1.5,~\kappa=0.5$, (c) $m_r=1.5,~n_r=0.8,~\kappa=8.5$ respectively, (d) Plot of eigen frequency and (e) propagation of nonlinear mode for $m_r=12.74\,,n_r=-9.01,n_i=-0.5,~u=0.0973,~\kappa=0.0049$.
	}
\end{figure}
\begin{figure}[]
	\centering
	\subfloat[\label{}]{\includegraphics[width=5cm,height=4cm]{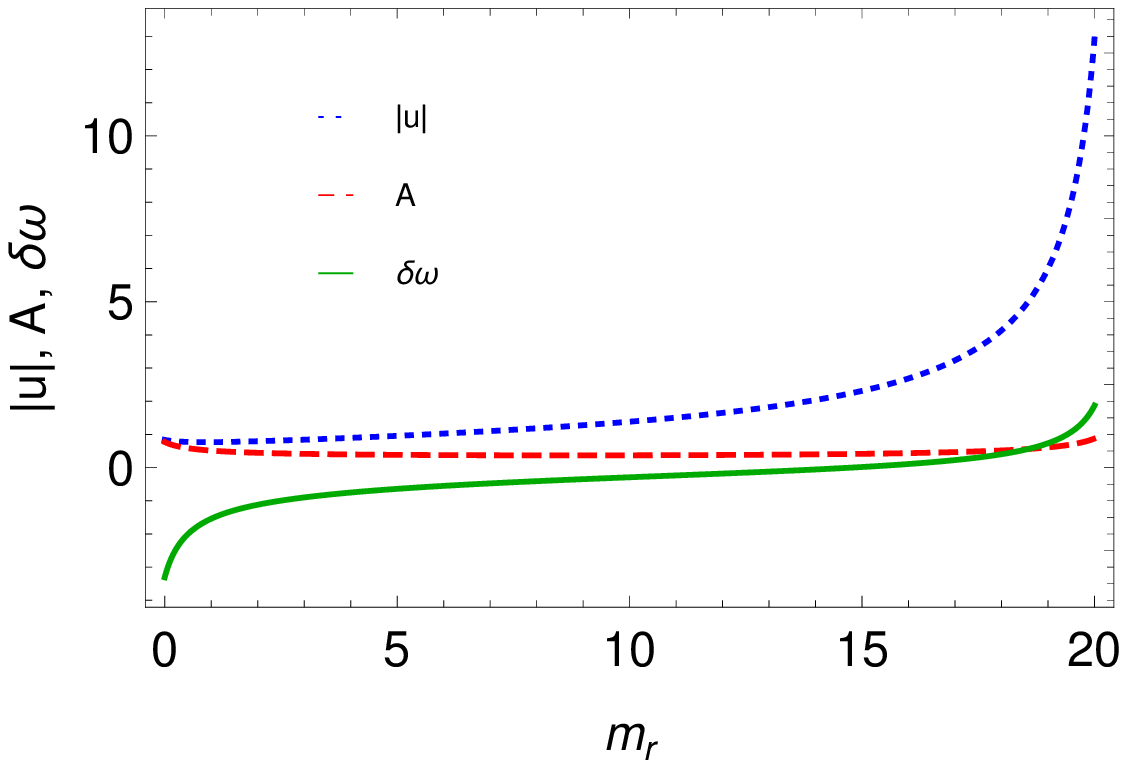}}
	~~
	\subfloat[\label{}]{\includegraphics[width=5cm,height=4cm]{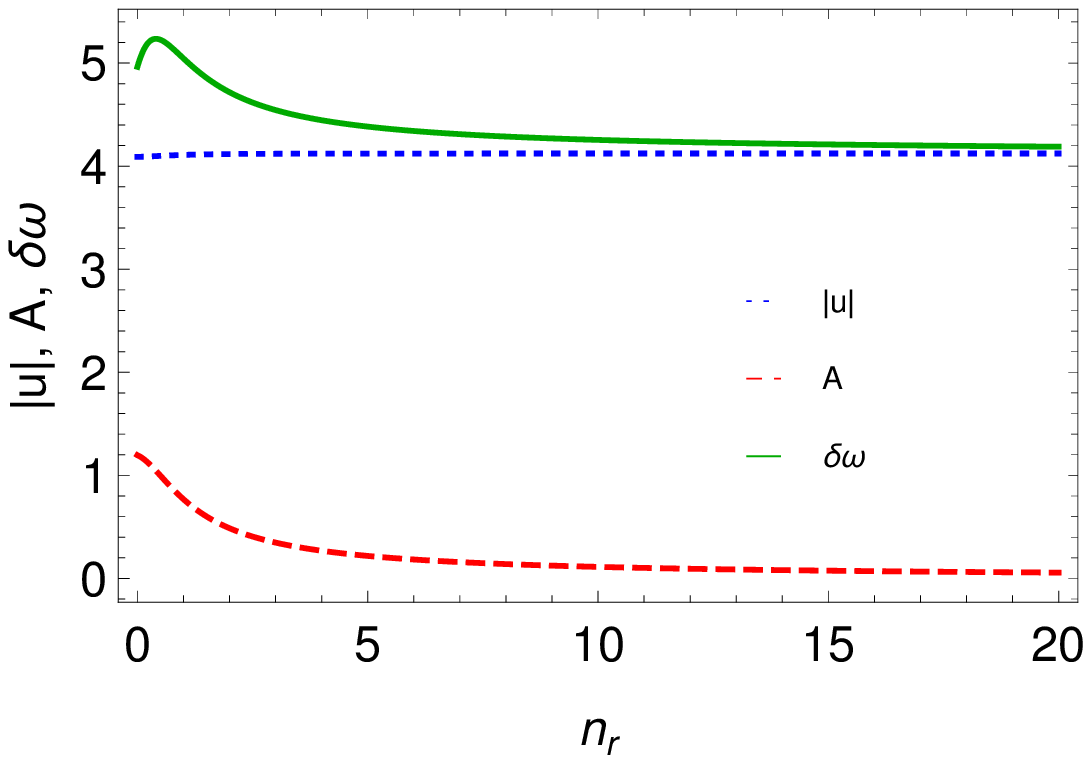}}
	~~
	\subfloat[\label{}]{\includegraphics[width=5cm,height=4cm]{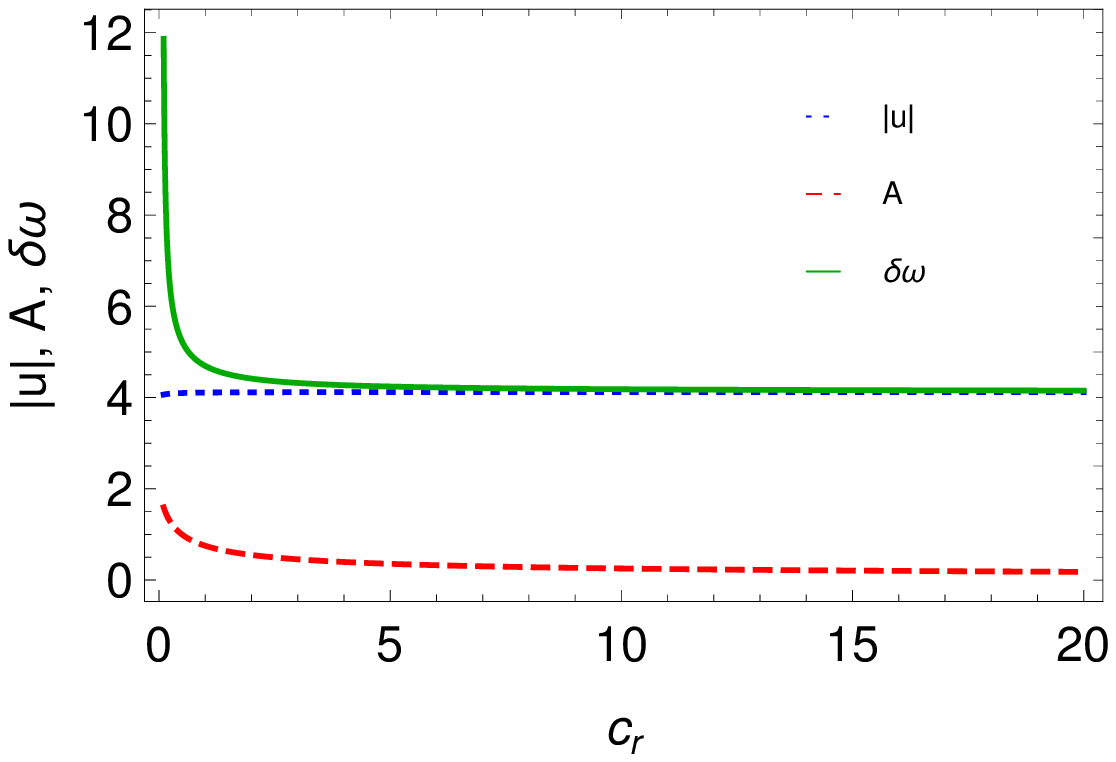}}\\
	~~
	~~
	\subfloat[\label{}]{\includegraphics[width=5cm,height=4cm]{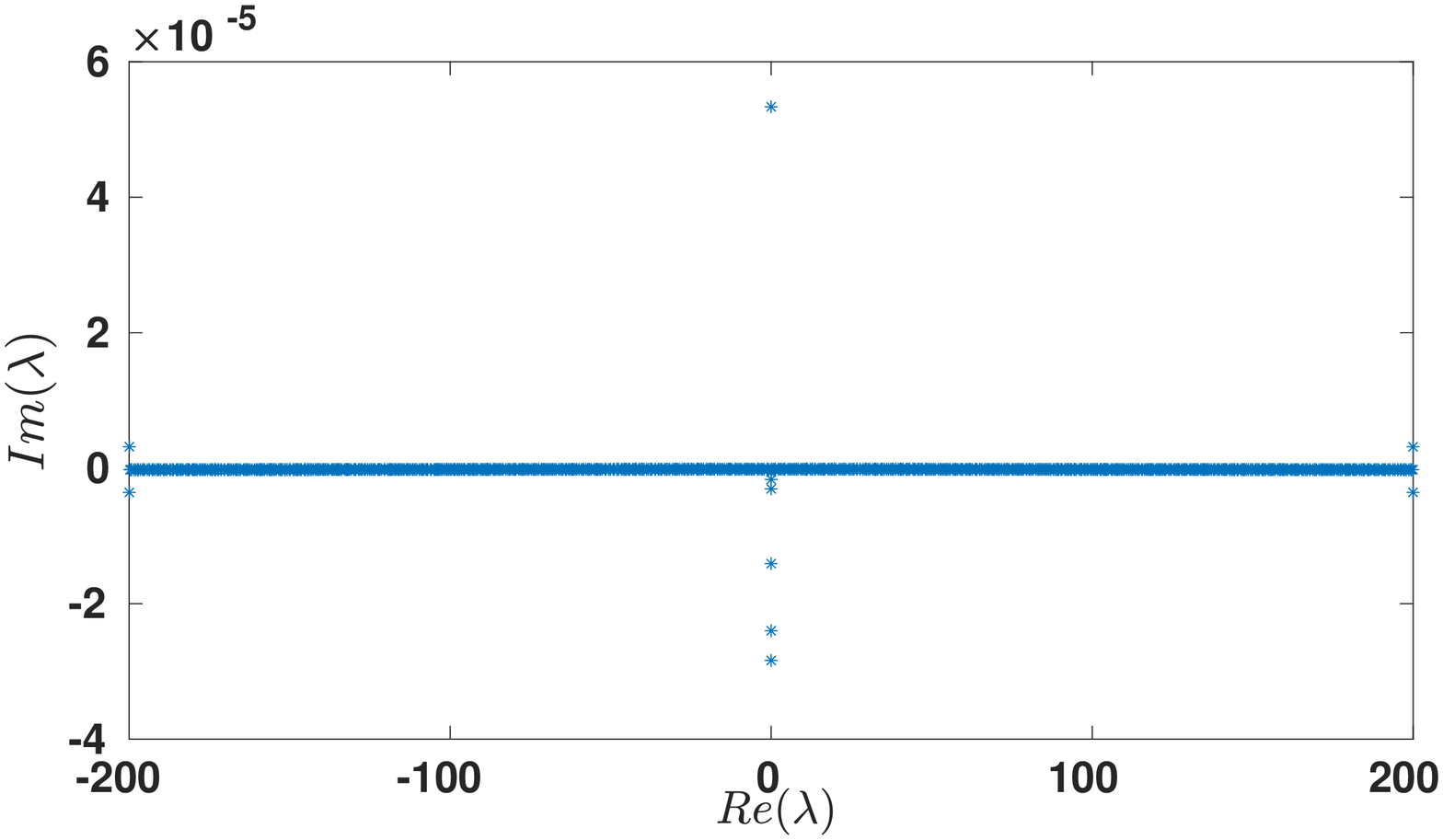}}
	~~
	\subfloat[\label{}]{\includegraphics[width=5cm,height=4cm]{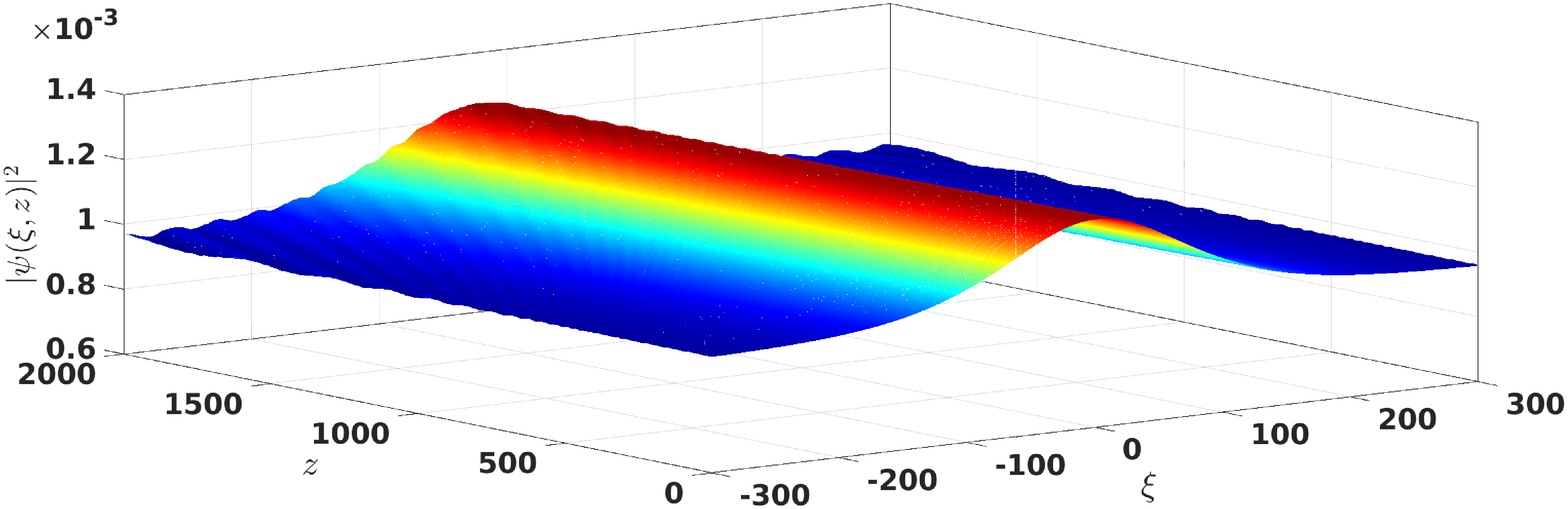}}\\
	
	\caption{For the solution(\ref{19}) with conditions (\ref{20}) and (\ref{20b}) taking $p_r=-1, D= -0.3$ (a), (b), (c) are plots of $|u|$ (blue dotted line), $A$ (red dashed line), $\delta\omega$ (green solid line)
		vs $m_r,~n_r,~c_r$ respectively at  $\xi=5,~\xi_0=0$. Other parameters are $n_r=-4.9,~c_r=-13,~\kappa=0.1$ in (a), $m_r=0.3,~c_r=0.5,~\kappa=8.5$ in (b), $m_r=0.3,~n_r=0.5,~\kappa=8.5$ in (c) respectively. (d) Plot of eigen frequency and (e) propagation of nonlinear mode for $m_r=4.74\,,n_r=-2.01,n_i=0.05,~u=0.0021,~\kappa=-0.000497$.
	}
	
\end{figure}
\section{Linear Stability Analysis}
For checking the linear stability of the nonlinear modes $\psi(z,t)$, the perturbation is taken as 				
\begin{equation}\label{21}
	\psi(z,t)=[\rho(\xi)+\epsilon\{F_\lambda(\xi) e^{i\lambda z}+G_\lambda^*(\xi) e^{-i\lambda^* z}\}]e^{i[\chi(\xi)-\kappa z]}
\end{equation}
where $\epsilon<<1$ and $F_\lambda(\xi)$ and $G_\lambda(\xi)$  are the perturbation eigenfunctions corresponding to eigen frequency $\lambda$. Substituting (\ref{21}) into
Eqn.(\ref{1}) and then linearizing with respect to $\epsilon$, we obtain the
eigenvalue problem as
\begin{equation}\label{22}
	\left(\begin{array}{cc}
		L_{1} &  L_{2} \\
		-L_2^*&  -L_1^* \\
	\end{array}\right)
	\left(\begin{array}{c}
		F_\lambda(\xi)\\
		G_\lambda(\xi)\\
	\end{array}\right)
	=\lambda\left(\begin{array}{c}
		F_\lambda(\xi)\\
		G_\lambda(\xi)\\
	\end{array}\right),
\end{equation}
where
\begin{gather*}
	\begin{split}
		L_1 = \frac{p_r}{2}\frac{d^2}{d\xi^2}+[-iu+ip_r(\alpha \rho^2+\beta)-2i(m_r+im_i)\rho^2
		-i(n_r+in_i)\rho^2]\frac{d}{d\xi}+u\alpha\rho^2\\+u\beta+\kappa+\frac{p_r}{2}(2i\alpha\rho\rho^{'}-\alpha^2\rho^4-2\alpha\beta\rho^2-\beta^2)+2(1+iq_i)\rho^2+3(c_r+ic_i)\rho^4\\
		-i(m_r+im_i)[4\rho\rho^{'}+2i(\alpha\rho^2+\beta)\rho^2]-3i(n_r+in_i)\rho\rho^{'},
	\end{split}\\
	\begin{split}
		L_2 =-i[(m_r+n_r)+i(m_i+n_i)]\rho^2\frac{d}{d\xi}+(1+iq_i)\rho^2+2(c_r+ic_i)\rho^4\\-i(m_r+im_i)[2\rho\rho^{'}+i(\alpha\rho^2+\beta)\rho^2]-i(n_r+in_i)\rho\rho^{'}.
	\end{split}
\end{gather*}

The nonlinear wave is propagating in the positive $z$ direction. So,  
for  $\lambda = \lambda_r + i\lambda_i$, (\ref{21}) shows that the nonlinear mode will have stable propagation  if $\lambda_i$ is nonnegative.
The linear stability analysis has been corroborated by direct numerical simulation adding $10\%$ random noise to the exact solutions. 

The eigen frequency spectra for the 
bright (\ref{11}), dark (\ref{16}), kink (\ref{14}), antikink (\ref{14a}), grey
(\ref{19}) with conditions (\ref{20a}) and antidark (\ref{19}) with conditions 
(\ref{20b}) are depicted in Figs. 1(d), 2(d), 3(d), 4(d), 5(d), 6(d) 
respectively. It is seen that for the chosen parameter values, the eigenfrequency spectrum of all the solutions has negative imaginary part. This is why the instability in the propagation of all the nonlinear modes is seen after having stable propagation upto approximately 1000 units.

\section{Summary:}
The complex cubic quintic Ginzburg Landau equation in 
the presence of the nonlinear gradient terms like SS, SFS and the nonlinear gain/loss is considered. This system models localized state
formation in mode locked lasers \cite{schalte}. A potentially rich set of exact
solitary waves such as, bright, dark, antidark, grey, kink, antikink (obeying certain parametric conditions) are
derived. 
It is found that chirping of all the nonlinear modes can be controlled by varying self steepening, delayed nonlinear gain, and quintic Kerr effect. 
The novelty of the present formulation lies in the existence of the 
grey, antidark, antikink dissipative solitary waves.The stability 
of the obtained solitary waves are investigated by linear stability analysis and the results are corroborated by direct numerical simulations. The solitary waves exhibit stable evolution up to a distance of approximately $1000$ units along the direction of propagation for the chosen parameter values. The effect of the variation of the model parameters (independent) 
on the physical quantities such as, the amplitude, speed and the chirping are 
explored. The results presented here may find applications in ultrashort pulse 
propagation in optical transmission line, in the design of optical fiber 
amplifiers, optical pulse compressors, dynamics of multimode lasers, and 
parametric oscillators \cite{schalte}. For future study, it would be interesting to 
obtain chirped dissipative solitary waves in the presence of higher order 
quintic, septimal gradient terms and the associated nonlinear gain/loss. This 
has relevance in obtaining high repetition rate (e.g. beyond ultrashort, 
attosecond) pulse sources based on the fiber technology \cite{hadrich}.

\ed
\begin{thebibliography}{}
	\bibitem{akhme1} N.Akhmediev and A.Ankiewicz Dissipative Solitons : Lecture Notes in Physics, Vol.661, edited by N.Akhmediev and A.Ankiewicz (Springer, Berlin, 2005);  I.S.Aranson, L.Kramer, Rev.Mod.Phys. {\bf 74}, 99 (2002); N.Akhmediev and A.Ankiewicz, Dissipative Solitons: From Optics to Biology and Medicine, Lecture Notes in Physics (Springer,Berlin,2008).
	\bibitem{moor} J.D.Moor, Opt.Commun. {\bf 96}, 65 (1993).
	\bibitem{matsu} A.Mecozzi et al, Opt.Lett.{\bf 16},1841 (1991); Y.Kodama, A.Hasegawa, Opt.Lett. {\bf 17}, 31 (1992); L.F.Mollenauer et al, Opt.Lett. {\bf 17}, 1575(1992).
	\bibitem{haus} H.Haus et al, J.Opt.Soc.Am.B {\bf 8}, 2068 (1991); P.Grelu, N.Akhmediev, Nat.Photon.{\bf 6}, 84 (2012).
	\bibitem{facao} M.Facao et al, Phys.Rev.A {\bf 91}, 013828 (2015).
	\bibitem{akhme661} N.Akhmediev and A.Ankiewicz, Solitons: Nonlinear Pulses and Beams (Chapman and Hall, London, 1997). 
	\bibitem{crespo} J.M.Soto-Crespo et al, Phys.Rev.E {\bf 70}, 066612 (2004); L.Duan et al, Opt.Exp. {\bf 20}, 265 (2012); H.Chen et al, Opt.Lett. {\bf 41}, 4210 (2013); L.M.Zhao et al, Opt.Lett. {\bf 38}, 1942 (2013).
	\bibitem{marq} P. Mareq et al, Physica D {\bf 73}, 305 (1994);  S.Chen et al, Phys.Rev.A {\bf 81}, 061806(R) (2010);  N. N. Akhmediev et al, Phys. Rev. Lett. {\bf 79}, 4047 (1997); S.C.Mancas, S.Roy Choudhury, Non.Dyn. {\bf 79}, 549 (2015); O.Descalzi, H.R.Brand, Phys.Rev.E {\bf 81}, 026210 (2010);  O.Descalzi et al, Phys.Rev.E {\bf 67}, 015601(R) (2003);  D.Turaev et al, Phys.Rev.E {\bf 75}, 045601(R) (2007); B.A.Malomed, A.A.Nepomnyashchy, Phys.Rev.A {\bf 42}, 6009 (1990);  
	V.Skarka, N.B.Aleksic, Phys.Rev.Lett. {\bf 96}, 013903 (2006).   
	\bibitem{akhme20} N. Akhmediev et al, Phys. Rev. E {\bf 63}, 056602 (2001); J. M. Soto-Crespo et al, Phys. Rev. Lett. {\bf 85}, 2937; J.M.Soto-Crespo et al, Phys.Lett.A {\bf 291}, 115(2001).
	\bibitem{akhme10} V.L.Kalashnikov, Phys.Rev.E {\bf 80}, 046606 (2009).  
	N.N.Akhmediev et al, Phys.Rev.E {\bf 53}, 1190 (1996); J.M.Soto-Crespo et al, J.Opt.Soc.Am.B {\bf 13}, 1439 (1996); W. H. Renninger et al, Phys.Rev.A  {\bf 77}, 023814 (2008); J.M.Soto-Crespo et al, Phys.Rev.E {\bf 55}, 4783 (1997); J.M.Soto-Crespo, L.Pesquera, Phys.Rev.E {\bf 56}, 7288 (1997); S.Chen, Phys.Rev.E {\bf 78}, 025601(R) (2008); E.N.Tsoy et al, Phys.Rev.E {\bf 73}, 036621 (2006).
	\bibitem{deiss1990} R.J.Deissler, H.R.Brand, Phys.Lett.A {\bf 146}, 252 (1990).
	\bibitem{latas1} S.C.V.Latas et al, Appl.Phys.B {\bf 104}, 131 (2011); S.C.V.Latas, M.F.S.Ferreira, Opt.Lett. {\bf 37}, 3897 (2012), Opt.Lett. {\bf 35}, 1771 (2010); L. Song et al, Opt. Commun. {\bf 249}, 301 (2005);  H.Tian et al, Appl.Phys.B {\bf 78}, 199 (2004);	M.V.Facao, M.I.Carvalho, Phys.Lett.A {\bf 375}, 2327 (2011); M.I.Carvalho, M.Facao, Phys.Lett.A {\bf 376}, 950 (2012); M.Facao et al, Phys.Lett.A {\bf 374}, 4844 (2010).
	\bibitem{uzu1} I.M.Uzunov et al, Phys.Rev.E {\bf 97}, 052215 (2018); I.M.Uzunov et al, Phys.Rev.E {\bf 90}, 042906 (2014); S. C. V. Latas et al, J. Opt. Soc. Am. B {\bf 34}, 1033 (2017).
	\bibitem{schalte} S.V.Gurevich et al, Phys.Rev.A. {\bf 99}, 061803 (2019). 
	\bibitem{cartes2016} C.Cartes, O.Descalzi, Phys.Rev.A {\bf 93}, 031801(R) (2016).
	\bibitem{deiss1998} R.J.Deissler, H.R.Brand, Phys.Rev.Lett. {\bf 81}, 3856 (1998). 
	\bibitem{hori14} T.P.Horikis, M.J.Ablowitz, J.Opt.Soc.Am.B {\bf 31}, 2748 (2014).
	\bibitem{saka37} H.Sakaguchi et al, Opt.Lett. {\bf 43}, 2688 (2018).
	\bibitem{tiofack2009} C.G.L.Tiofack et al, Phys.Rev.E {\bf 80}, 066604 (2009).
	\bibitem{tian10} H.Tian et al, Phys.Rev.E {\bf 66}, 066204 (2002).
	\bibitem{kalash13} V.L.Kalashnikov et al, Opt.Exp. {\bf 16}, 4206 (2008).
	\bibitem{bednya} A.E.Bednyakova et al, Opt.Exp. {\bf 21}, 20556 (2013).
	\bibitem{kalash14} V.L.Kalashnikov,E.Sorokin, Opt.Exp. {\bf 22}, 30118 (2014).
	\bibitem{nisha2020} Nisha et al, Phys. Lett.A {\bf 384}, 126675 (2020).
	\bibitem{belan6} P.A.Belanger, Opt.Exp. {\bf 14}, 12174 (2006).
	\bibitem{tian2005} J.Tian et al, Phys.Scr. {\bf 71}, 507 (2005).
	\bibitem{agra} G. Agrawal, Nonlinear Fiber Optics, 5th ed. (Academic Press,
	Boston, 2012).
	\bibitem{lam} M. R. E. Lamont et al, Opt. Lett. {\bf 38},
	3478 (2013).
	\bibitem{carval87} M.I.Carvalho, M.Facao,Phys.Rev.E {\bf 100}, 032222 (2019).
	\bibitem{koma} A.Komarov et al, Phys.Rev.E {\bf 72}, 025604(R) (2005).
	\bibitem{behera} A.Khare et al, J. Phys. {\bf A42}, 475404 (2009).
	\bibitem{hadrich} S.Hadrich et al, Opt.Exp. {\bf 18}, 0242 (2010); Y.Song et al, Opt.Exp. {\bf 19}, 14518 (2011).
	
\end{thebibliography}
